\documentclass[journal]{IEEEtran}
\usepackage[english]{babel}
\usepackage[a4paper,top=2cm,bottom=2cm,left=1.5cm,right=1.5cm,marginparwidth=1.75cm]{geometry}
\usepackage{float}
\usepackage{orcidlink}
\usepackage{booktabs}
\usepackage{amsmath}
\usepackage{amsfonts}
\usepackage{amssymb}
\usepackage{algorithm}
\usepackage{algorithmic}
\usepackage{cite}
\usepackage{graphicx, subfigure}
\usepackage{url}
\usepackage{bm}
\usepackage{soul}
\title{Short-Time Variational Mode Decomposition}
\author{Hao Jia,
        Pengfei Cao,
        Tong Liang,
        Cesar F. Caiafa,
        Zhe Sun,
        Yasuhiro Kushihashi,
        Grau A,
        Bolea Y,
        Feng Duan,
        Jordi Sol{\'e}-Casals
}

\author{%
    \IEEEauthorblockN{%
        Hao Jia\IEEEauthorrefmark{1}\IEEEauthorrefmark{2}, 
        Pengfei Cao\IEEEauthorrefmark{3},
        Tong Liang\IEEEauthorrefmark{4},
        Cesar F. Caiafa\IEEEauthorrefmark{5},
        Zhe Sun\IEEEauthorrefmark{6},
        Yasuhiro Kushihashi\IEEEauthorrefmark{4},
        Antoni Grau\IEEEauthorrefmark{3},
        Yolanda Bolea\IEEEauthorrefmark{3},
        Feng Duan\IEEEauthorrefmark{1}\IEEEauthorrefmark{2},
        Jordi Solé-Casals\IEEEauthorrefmark{7}%
    }%
    
    \IEEEauthorblockA{%
        \IEEEauthorrefmark{1}School of Medicine, Nankai University, Tianjin, China\\
        \IEEEauthorrefmark{2}Tianjin Key Laboratory of Interventional Brain-Computer Interface and Intelligent Rehabilitation, Tianjin, China\\
        \IEEEauthorrefmark{3}Universitat Politècnica de Catalunya, Barcelona, Spain\\
        \IEEEauthorrefmark{4}Mechanical Systems Engineering, Nippon Institute of Technology, Saitama, Japan\\
        \IEEEauthorrefmark{5}Instituto Argentino de Radioastronomía (CCT-CONICET La Plata/CIC-PBA/UNLP), Villa Elisa, Argentina\\
        \IEEEauthorrefmark{6}Computational Bioengineering Laboratory, Faculty of Health Data Science, Juntendo University, Tokyo, Japan\\
        \IEEEauthorrefmark{7}Data and Signal Processing Research Group, University of Vic-Central University of Catalonia, Vic, Spain%
    }%
    \thanks{Corresponding authors: Jordi Solé-Casals (email: jordi.sole@uvic.cat) and Feng Duan (email: duanf@nankai.edu.cn)}
}

\begin{document}
\maketitle

\begin{abstract}
Variational mode decomposition (VMD) and its extensions like Multivariate VMD (MVMD) decompose signals into ensembles of band-limited modes with narrow central frequencies. These methods utilize Fourier transformations to shift signals between time and frequency domains. However, since Fourier transformations span the entire time-domain signal, they are suboptimal for non-stationary time series.

We introduce Short-Time Variational Mode Decomposition (STVMD), an innovative extension of the VMD algorithm that incorporates the Short-Time Fourier transform (STFT) to minimize the impact of local disturbances. STVMD segments signals into short time windows, converting these segments into the frequency domain. It then formulates a variational optimization problem to extract band-limited modes representing the windowed data. The optimization aims to minimize the sum of the bandwidths of these modes across the windowed data, extending the cost functions used in VMD and MVMD. Solutions are derived using the alternating direction method of multipliers, ensuring the extraction of modes with narrow bandwidths.

STVMD is divided into dynamic and non-dynamic types, depending on whether the central frequencies vary with time. Our experiments show that non-dynamic STVMD is comparable to VMD with properly sized time windows, while dynamic STVMD better accommodates non-stationary signals, evidenced by reduced mode function errors and tracking of dynamic central frequencies. This effectiveness is validated by steady-state visual-evoked potentials in electroencephalogram signals.
\end{abstract}

\section{Introduction}
\IEEEPARstart{N}{on-stationary} signals, characterized by time-varying statistical properties like frequency, are prevalent in fields like biomedical engineering, telecommunications, and environmental monitoring. Signal processing methods on non-stationary signals have attracted a lot of interests due to its wide applications in the real-world. Here we focus on tracking the time variations of packets of close frequencies in input data, i.e., decomposing the input data into multiple packets with fewer overlapped frequency bands.

Empirical mode decomposition (EMD) is a method of analysing non-stationary signals based on the concept of adaptive decomposition of a signal into mode functions \cite{huang_empirical_1998}. EMD decomposes a signal into a finite number of so-called intrinsic mode functions (IMFs), each of which represents a locally defined oscillatory mode with a characteristic frequency and time scale. This process is solved with a recursive sifting process on the local extrema. In the sifting process, the local extrema of signals are interpolated to obtain the upper and lower envelopes. The local mean, obtained by averaging the upper and lower envelopes, is regarded as the low-frequency estimate of the data. The local mean is removed from the data recursively to yield the mode functions with high frequencies. The sifting process stops until the stop criterion reaches, e.g., there is no enough extremes to construct the envelopes.

The multivariate EMD (MEMD), as an extension of EMD algorithm, also make use of signal extrema to construct the envelopes \cite{rehman_multivariate_2010}. MEMD considers the decomposition of multiple time series simultaneously. Instead of calculating the signal extrema of each time series independently, MEMD casts the signals onto a sphere surface and locates the local extrema on the sphere surface. The locations of the extrema are used to construct the envelopes in the time domain.

Unlike traditional Fourier-based methods, EMD-based algorithms do not require predefined basis functions and adapt to the data's local characteristics, making them particularly suitable for nonlinear and non-stationary signals. A recent advancement in this direction is the difference mode decomposition \cite{hou2023difference}, which provides enhanced adaptive signal decomposition capabilities. However, EMD-based algorithms may suffer from mode mixing, where the extracted mode functions contain components from multiple frequency scales, necessitating further refinements and enhancements to improve their performance. Besides, because of the loss of mathematical theory, it is difficult to obtain theoretical guarantee for EMD-based algorithms.

Variational mode decomposition (VMD) is an analytic method that addresses some of the limitations of EMD \cite{dragomiretskiy_variational_2014}. VMD decomposes a single time series signal into multiple mode functions by solving a constrained variational problem. Each mode function is assumed to be mostly narrow-band around a central frequency, and the aim is to determine these modes and their corresponding central frequencies. Recent work has expanded VMD's capabilities through variational time-frequency adaptive decomposition \cite{zhao2023variational}, improved spectrum reconstruction techniques \cite{meng2023improved}, and adaptive parameter selection methods \cite{yu2023strain}.

In VMD, the frequency content of the signals, generated by the Fourier transform, is decomposed into a finite number of modes by solving a convex optimization problem. The objective of the optimization problem is to minimize the collective bandwidth of mode functions, subjecting the constraint that the input signals can be recovered from the mode functions. The resulting objective function is minimized using the alternating direction method of multipliers (ADMM) to find an ensemble of modes and the central frequencies of these modes. Recent advancements include generalized nonlinear mode decomposition \cite{wang2024variational} and efficient algorithms for specific applications like lung sound analysis \cite{liu2024efficient}.

While VMD handles univariate signals, many real-world applications involve multivariate signals. The multivariate signals refer to the multiple time series which are supposed to be processed simultaneously. To address this need of multivariate signal processing, Multivariate variational mode decomposition (MVMD) extends the VMD framework to multivariate signals \cite{rehman_multivariate_2019}. MVMD jointly decomposes the multivariate signals into a set of common modes, ensuring that the consistent modes across all time series and capturing the underlying correlations between them. Recent developments include grouped MVMD for EEG analysis \cite{jian_grouped_2024}, adaptive multivariate chirp mode decomposition \cite{huang2023adaptive}, and optimized VMD algorithms for biomedical signal processing \cite{chen2024optimized}.

In the EMD-based decomposition method, the decomposition process is related to the local extremes, and central frequencies of decomposed mode functions are not expressed as a numerical value during the sifting process. In the VMD-based decomposition method, the central frequencies are estimated and the mode functions shares the same central frequencies as time changes. Some extensions of VMD and EMD improved the decomposition performance, such as the noise-assisted EMD \cite{yeh_complementary_2010}, noise-assisted VMD \cite{zisou_noise-assisted_2021}, or ensemble EMD \cite{wu_ensemble_2009}. Target at solving the limitation of predefined number of mode functions, Successive VMD is proposed \cite{liu_successive_2022}. Some extensions, like grouped MVMD, are proposed to solve the real-world applications \cite{jian_grouped_2024}.

In the stationary signal, the central frequencies of signals are fixed and it is not necessary to track the dynamic changes of the central frequencies. Therefore, the EMD-based and VMD-based methods ignore the information provided by the central frequencies. In the non-stationary signals, the frequencies are usually time-varying. Tracking the dynamic changes of the central frequencies provides the decomposed methods more information.

In this work, we present a novel extension of the VMD algorithm, named short-time VMD (STVMD), to process multivariate signals with time-varying frequencies. Our proposed STVMD is of two types, non-dynamic STVMD and dynamic STVMD. In the non-dynamic STVMD, the central frequencies are fixed as time changes, exhibiting the same properties as VMD and MVMD with a properly selected length of time windows. In the dynamic STVMD, the central frequencies are time-varying, making STVMD more adaptive to non-stationary signals.

The proposed STVMD model is developed by replacing the Fourier transform with the short-time Fourier transform (STFT) in VMD, avoiding the negative influence of signal changes in the global time domain. In the STVMD model, we seek a collection of common modulated short-time mode functions residing in the input signals that exhibit a minimum collective bandwidth while fully reconstructing the input signals. The key innovation of the proposed method is summarised in the following three points:

\noindent (1) The mode decomposition methods are extended to the field of non-stationary signal processing;

\noindent (2) The central frequencies of decomposed mode functions exhibit the dynamic changes of frequencies in the non-stationary signals with time-varying frequencies;

\noindent (3) STVMD is not only a decomposition method but also a signal analysis approach.

The paper is organized as follows. Section \ref{sec:2} and \ref{sec:3} review the basic concepts for VMD and STFT, respectively. Section \ref{sec:4} explains the non-dynamic STVMD and dynamic STVMD in detail. Section \ref{sec:5} systematically analyzes the relationship between VMD, non-dynamic STVMD and dynamic STVMD.

\section{Variational Mode Decomposition}
\label{sec:2}
Variational Mode Decomposition (VMD) is a sophisticated signal processing technique designed to decompose a given signal into a set of modes, each with a specific central frequency and narrow bandwidth. Providing a robust and mathematically principled framework, VMD can be divided into several key components: AM-FM modulation, Hilbert transform, and the variational decomposition process itself.

\subsection{AM-FM Modulation}

In signal processing, a signal can often be represented as a combination of Amplitude Modulation (AM) and Frequency Modulation (FM) components. Mathematically, an AM-FM signal can be expressed as

\begin{equation}
u(t) = a(t) \cos(\phi(t)),
\end{equation}
where \( a(t) \) represents the instantaneous amplitude, and \( \phi(t) \) is the instantaneous phase. The instantaneous frequency is then given by the derivative of the phase:

\begin{equation}
\omega(t) = \frac{d\phi(t)}{dt}.
\end{equation}

VMD aims to decompose a signal into a finite number of such AM-FM components, each characterized by a specific central frequency and a narrow bandwidth.

\subsection{Hilbert Transform}

The Hilbert transform is a tool to obtain the analytic signal from a real-valued signal in VMD. The analytic signal \( u_+(t) \) of a real-valued signal \( u(t) \) is defined as:

\begin{align*}
    u_+(t) & = u(t) + j \mathcal{H}\{u(t)\} \\
    & = a(t)(\cos(\phi(t))+j\sin(\phi(t)))\\
    & = a(t)e^{j\phi(t)},
\end{align*}
where \( \mathcal{H}\{*\} \) denotes the Hilbert transform. The original (real) signal is retrieved as the real part of the analytic signal:
\begin{equation}
    u(t) = \mathcal{R}\{u_+(t)\}.
\end{equation}

\subsection{Variational Mode Decomposition}

The VMD process involves decomposing a signal into \( K \) modes, each with an associated central frequency. This is achieved through the following steps:

\subsubsection{Initialization}

- The number of modes \( K \) to be extracted is predetermined.
- Initialize the modes \( u_k \) and their associated central frequencies \( \omega_k \).

\subsubsection{Formulation of the Variational Problem}

The VMD method formulates the decomposition as a constrained variational problem. The objective is to minimize the sum of the bandwidths of the modes while ensuring that their sum reconstructs the original signal $x(t)$. The variational problem is defined as:

\begin{equation}
\min_{\{u_k\}, \{\omega_k\}} \left\{ \sum_{k=1}^K \left\| \partial_t \left[u_+^k(t) e^{-j \omega_k t} \right] \right\|_2^2 \right\},
\end{equation}
subject to:
\begin{equation}
\sum_{k=1}^K u_k(t) = x(t),
\end{equation}
where $u_+^k(t)$ is the analytic signal of the decomposed mode functions $u_k(t)$, and \( \omega_k \) are their central frequencies. The term \( \partial_t \) denotes the partial derivative with respect to time.

\subsubsection{Augmented Lagrangian Form}

To solve the constrained optimization problem, the augmented Lagrangian is constructed as follows:
\begin{align}
\mathcal{L}(\{u_k\}, &\{\omega_k\}, \lambda) = \alpha\sum_{k=1}^K \left\| \partial_t \left[u_+^k(t) e^{-j \omega_k t} \right] \right\|_2^2 \nonumber \\
    & + \left\| x(t) - \sum_{k=1}^K u_k(t) \right\|_2^2 + \left< \lambda(t), x(t) - \sum_{k=1}^K u_k(t) \right>,
\end{align}
where \( \lambda(t) \) is the Lagrange multiplier. The augmented term \( \left\| x(t) - \sum_{k=1}^K u_k(t) \right\|_2^2 \) ensures that the reconstruction constraint is met.

\subsubsection{Alternating Direction Method of Multipliers (ADMM)}

To solve the variational problem, the Alternating Direction Method of Multipliers (ADMM) is used. This iterative optimization breaks down the problem into simpler subproblems:

\noindent(i) {Update of Modes \( u_k \)}

Each mode \( u_k \) is updated by solving:

\begin{align}
u_k^{n+1}(t) =& \arg \min_{u_k} \left\{ \alpha\left\| \partial_t \left[u_+^k(t) e^{-j \omega_k^n t} \right] \right\|_2^2 \right. \nonumber \\
&\quad \left. + \left\| x(t) - \sum_{i=1}^K u_i(t)^n+\frac{\lambda^n(t)}{2} \right\|_2^2 \right\}.
\end{align}

\noindent(ii) {Update of Frequencies \( \omega_k \)}

Each central frequency \( \omega_k \) is updated using:

\begin{equation}
\omega_k^{n+1} = \frac{\int_0^\infty t |u_k^{n+1}(t)|^2 \, dt}{\int_0^\infty |u_k^{n+1}(t)|^2 \, dt}.
\end{equation}

\noindent(iii) {Update of Lagrange Multipliers}

The Lagrange multipliers are updated to enforce the reconstruction constraint:

\begin{equation}
\lambda^{n+1}(t) = \lambda^n(t) + \tau \left( x(t) - \sum_{k=1}^K u_k^{n+1}(t) \right),
\end{equation}
where \( \tau \) is a step size parameter.

Iterative updates continue until the convergence criteria are met, typically defined by the error between $x$ and $\sum_{i=1}^K u_i^n$ falling below a predefined threshold.

\subsubsection{Solving Process}
The Parseval/Plancherel theorem states that the total energy (norm) of a signal is preserved under its Fourier transform. This property is crucial for spectral domain operations, ensuring that energy-conserving transformations lead to accurate reconstructions in the time domain. With this theorem, the update of modes $u_k$ is given by
\begin{equation}
    \hat{u}_k^{n+1}(\omega)=\frac{\hat{x}-\sum_{i!=k}\hat{u}_k(\omega)+\frac{\hat{\lambda(\omega)}}{2}}{1+2\alpha(\omega-\omega_k)^2},
    \label{eqn:uk}
\end{equation}
where $\hat{x}$ and $\hat{u}$ correspond to the $x(t)$ and $u(t)$ in the frequency domain, converted with Fourier transform. Meanwhile, the update of the frequencies $\omega_k$ is adapted to
\begin{equation}
    \omega_k^{n+1} \leftarrow \frac{\int_0^\infty \omega \left| \hat{u}_k^{n+1}(\omega) \right|^2 d\omega}{\int_0^\infty \left| \hat{u}_k^{n+1}(\omega) \right|^2 d\omega}.
    \label{eqn:wk}
\end{equation}
The optimization process of VMD is given in Algorithm \ref{alg:2}, in the case that the number of channels $C$ equals to 1. More details about the VMD algorithm can be found in \cite{dragomiretskiy_variational_2014}.

% \begin{algorithm}[htbp]
% \caption{Complete optimization of VMD}
% \begin{algorithmic}
% \STATE Initialize \( \{\hat{u}_k^1\}, \{\omega_k^1\}, \hat{\lambda}^1, n \leftarrow 0, \epsilon \leftarrow 1e^{-9}\)
% \REPEAT
%     \STATE \( n \leftarrow n + 1 \)
%     \FOR{ $k = 1 : K$ }
%         \STATE \textbf{Update \( \hat{u}_k \) for all \( \omega \geq 0 \):}
%         \STATE
%         $\hat{u}_k^{n+1}(\omega) \leftarrow$
%         \begin{equation}
%             \frac{\hat{x}(\omega) - \sum_{i<k} \hat{u}_i^{n+1}(\omega) - \sum_{i>k} \hat{u}_i^n(\omega) + \frac{\hat{\lambda}^n(\omega)}{2}}{1 + 2\alpha (\omega - \omega_k^n)^2}
%             \nonumber
%         \end{equation}
%     \ENDFOR
%     \FOR{ $k = 1 : K$ }
%         \STATE \textbf{Update \( \omega_k \):}
%         \begin{equation}
%         \omega_k^{n+1} \leftarrow \frac{\int_0^\infty \omega \left| \hat{u}_k^{n+1}(\omega) \right|^2 d\omega}{\int_0^\infty \left| \hat{u}_k^{n+1}(\omega) \right|^2 d\omega}
%         \nonumber
%         \end{equation}
%     \ENDFOR
%     \STATE \textbf{Dual ascent for all \( \omega \geq 0 \):}
%     \begin{equation}
%     \hat{\lambda}^{n+1}(\omega) \leftarrow \hat{\lambda}^n(\omega) + \tau \left( \hat{x}(\omega) - \sum_{k} \hat{u}_k^{n+1}(\omega) \right)
%     \end{equation}
% \UNTIL{convergence: \( \sum_k \frac{\|\hat{u}_k^{n+1} - \hat{u}_k^n\|_2^2}{\|\hat{u}_k^n\|_2^2} < \epsilon \).}
% \end{algorithmic}
% \label{alg:1}
% \end{algorithm}

\subsection{Multivariate Variational Mode Decomposition}
Multivariate Variational Mode Decomposition (MVMD) is an extension of the VMD algorithm. In VMD, a single time series is decomposed into multiple mode functions. In biomedical signal processing, the multivariate time series are often acquired and analysed, which encouraged the proposal of MVMD. MVMD considers the decomposition of multiple time series simultanously.

\begin{algorithm}[htbp]
\caption{Complete optimization of (M)VMD}
\begin{algorithmic}
\STATE Initialize \( \{\hat{u}_{k,c}^1\}, \{\omega_k^1\}, \hat{\lambda}_c^1, n \leftarrow 0, \epsilon \leftarrow 1e^{-9}\)
\REPEAT
    \STATE \( n \leftarrow n + 1 \)
    \FOR{ $k = 1 : K$ }
    \FOR{ $c = 1 : C$ }
        \STATE \textbf{Update \( \hat{u}_{k,c} \) for all \( \omega \geq 0 \):}
        \STATE
        $\hat{u}_{k,c}^{n+1}(\omega) \leftarrow$
        \begin{equation}
            \frac{\hat{x_c}(\omega) - \sum_{i<k} \hat{u}_{i,c}^{n+1}(\omega) - \sum_{i>k} \hat{u}_{i,c}^n(\omega) + \frac{\hat{\lambda}_c^n(\omega)}{2}}{1 + 2\alpha (\omega - \omega_k^n)^2}
            \nonumber
        \end{equation}
    \ENDFOR
    \ENDFOR
    \FOR{ $k = 1 : K$ }
        \STATE \textbf{Update \( \omega_k \):}
        \begin{equation}
        \omega_k^{n+1} \leftarrow \frac{\sum_{c=1}^C\int_0^\infty \omega \left| \hat{u}_k^{n+1}(\omega) \right|^2 d\omega}{\sum_{c=1}^C\int_0^\infty \left| \hat{u}_k^{n+1}(\omega) \right|^2 d\omega}
        \nonumber
        \end{equation}
    \ENDFOR
    \FOR{ $c = 1 : C$ }
        \STATE \textbf{Update \( \hat{\lambda}_{c} \):}
        \begin{equation}
        \hat{\lambda}^{n+1}_c(\omega) \leftarrow \hat{\lambda}^n_c(\omega) + \tau \left( \hat{x}_c(\omega) - \sum_{k} \hat{u}_{k,c}^{n+1}(\omega) \right)
        \end{equation}
    \ENDFOR
\UNTIL{convergence: \( \max \frac{\|\hat{u}_{k,c}^{n+1} - \hat{u}_{k,c}^n\|_2^2}{\|\hat{u}_{k,c}^n\|_2^2} < \epsilon, c=1,2,\cdots,C\).}
\end{algorithmic}
\label{alg:2}
\end{algorithm}

In the decomposition of multiple time series, a simple approach is to decompose the time series with VMD one by one. In this case, the central frequencies $\omega_k$ of each time series are not synchronized so that the decomposed mode functions have different central frequencies across the time series.

In MVMD, this problem is solved by assuming that the central frequencies are the same among the time series. With this assumption, the augmented Lagrange form of MVMD is given by
\begin{align}
\mathcal{L}(\{u_k\}, &\{\omega_k\}, \lambda) = \alpha\sum_{c=1}^C\sum_{k=1}^K \left\| \partial_t \left[u_+^{k,c}(t) e^{-j \omega_k t} \right] \right\|_2^2 \nonumber \\
    & + \sum_{c=1}^C\left\| x_c(t) - \sum_{k=1}^K u_{k,c}(t) \right\|_2^2 \nonumber \\
    & + \sum_{c=1}^C\left< \lambda(t), x_c(t) - \sum_{k=1}^K u_{k,c}(t) \right>,
\end{align}
where \( \lambda(t) \) is the Lagrange multiplier; $C$ is the number of time series in the multivariate time series; $x_c$, $u_{k,c}$ and $u_+^{k,c}$ are the $x$, $u_{k}$ and $u_+^{k}$ of the $c$-th time series. In this objective function with Lagrange form, the $u_+^{k,c}$ shares the same $\omega_k$ for $c=1,2,\cdots,C$, which force the alignment of central frequencies of multiple time series. This objective function can be solved in the same approach as in VMD. The complete optimization process of MVMD is given in Algorithm \ref{alg:2}, and details about the MVMD algorithm can be found in \cite{rehman_multivariate_2019}.

\section{Short-time Fourier Transform}
\label{sec:3}
In the VMD process, the minimization of the augmented Largrangian function is solved in the frequency domain based on the Parseval/Plancherel theorem. The input signal $x(t)$ in the time domain is converted to $\hat{x}(\omega)$ frequency domain with Fourier transform, and the mode functions $\{\hat{u}_k(\omega)\}$ are reversed to $\{u_k(t)\}$ in the time domain.

The VMD algorithm is generalized into two points, time-frequency (T-F) transform and single-multiple (S-M) transform, as follows:

\begin{itemize}
    \item (T-F transform) convert the signal from the time domain to the frequency domain and its reverse;
    \item (S-M transform) decomposes the signal into multiple narrow bands in the frequency domain.
\end{itemize}

In the VMD algorithm and its extensions, like MVMD, the design of algorithms focuses on the S-M transform, but the influence of the T-F transform was not discussed. In the previous implementations of VMD and MVMD, the Fourier transform and its inverse are used in the T-F transform. 

Given an original time signal \( x(t) \), the Fourier transform \( F(\omega) \) of \( x(t) \) is defined by the integral:
\begin{equation}
    \hat{x}(\omega) = \mathcal{F}\{x(t)\} = \int_{-\infty}^{\infty} x(t) e^{-j \omega t} \, dt,
\end{equation}
where \( \omega \) is the angular frequency in radians per second, and \( j \) is the imaginary unit, \( j^2 = -1 \).
The inverse Fourier transform, which reconstructs the original time-domain signal from its frequency-domain representation, is given by:
\begin{equation}
    x(t) = \mathcal{F}^{-1}\{\hat{x}(\omega)\} = \frac{1}{2\pi} \int_{-\infty}^{\infty} \hat{x}(\omega) e^{j \omega t} \, d\omega.
\end{equation}

Since the Fourier transform is the integral of the entire signal in the time domain, any transient characteristics, such as noise spike, line update or abrupt signal change, are spread over the entire frequency spectrum. This spreading can obscure the true underlying behaviour of the signal and make it difficult to identify or isolate specific frequency components that are only present for short durations.

Compared to the Fourier transform, the STFT applies the integral to the windowed time-domain signal. Given the original time signal \( x(t) \) and the window function \( w(t) \),  the STFT \( \hat{x}(\tau, \omega) \) of \( x(t) \) is defined as:
\begin{equation}
    \hat{x}(\tau, \omega) = \int_{-\infty}^{\infty} x(t) w(t - \tau) e^{-j \omega t} \, dt,
\end{equation}
where \( \tau \) is the time index of the window's position. The window function \( w(t-\tau) \) is centered at \( t = \tau \) and is usually a smooth, decaying function.

The inverse STFT, which reconstructs the original signal \( x(t) \) from its STFT, is given by:
\begin{equation}
    x(t) = \frac{1}{2\pi} \int_{-\infty}^{\infty} \int_{-\infty}^{\infty} \hat{x}(\tau, \omega) w(t - \tau) e^{j \omega t} \, d\omega \, d\tau.
\end{equation}

Given discrete signals $x[n]\in\mathbb{R}^{1\times T}$, its Fourier transform has the format of 
\begin{equation}
    \hat{x}[m] = \mathcal{F}\{x[n]\}= \sum_{n=0}^{T-1} x[n] e^{-j 2 \pi m n / T}.
    \label{eqn:dft}
\end{equation}
The inverse discrete Fourier transform is given by
\begin{equation}
    x[n] = \mathcal{F}^{-1}\{\hat{x}[m]\}= \frac{1}{T}\sum_{n=0}^{T-1} \hat{x}[k] e^{j 2 \pi m n / T}.
    \label{eqn:idft}
\end{equation}

In the discrete STFT, a sliding time window $w[n]$ is applied to $x[n]$ and we have
\begin{equation}
    \bm{x}[n]=\begin{bmatrix}
                x[n]w[n-1] \\
                x[n]w[n-2] \\
                \vdots \\
                x[n]w[n-T] 
              \end{bmatrix}
              =\begin{bmatrix}
                xw[1] \\
                xw[2] \\
                \vdots \\
                xw[T] 
              \end{bmatrix},
    \label{eqn:xw}
\end{equation}
where \( w[n] \) is the window function of length \( N \) and $xw[\tau]$ denotes $x[n]w[n-\tau], \tau=1,2,...,T$. The step length between two windows is set to 1 considering the signals need to be recovered to time domain with inverse STFT.
\begin{equation}
w[n] =
    \begin{cases}
        w(n) & \text{if } -\frac{N}{2} < n \leq \frac{N}{2}; \\
        0 & \text{else}.
    \end{cases} 
\end{equation}
Reflective padding is applied to the input data $x[n]$ around the borders so that the output data $\bm{x}[n]$ has the same length as the input data after applying the windowing operation.
Ignoring the entries filled with zeros induced by zeros in the window function, $\bm{x}[n]\in \mathbb{R}^{N\times T}$. Associated with Equation \ref{eqn:dft} and \ref{eqn:idft}, the Fourier transform of $\bm{x}[n]$ is 
\begin{equation}
    \hat{\bm{x}}[m] = \mathcal{F}\{\bm{x}[n]\} = \sum_{n=0}^{T} \bm{x}[n] e^{-j 2 \pi m n / T},
\end{equation}
and its inverse Fourier transform is
\begin{equation}
    \bm{x}[n] = \mathcal{F}^{-1}\{\hat{\bm{x}}[m]\} = \frac{1}{T}\sum_{n=0}^{T-1} \hat{\bm{x}}[m] e^{j 2 \pi m n / T}.
\end{equation}
Finally, the signal $x[n]$ is recovered from the windowed signal $\bm{x}[n]$ with the known window:
\begin{equation}
    x[n]=\frac{\sum_{\tau=1}^{T}x[n]w[n-\tau]}{\sum_{\tau=1}^{T}1[n]w[n-\tau]},
    \label{eqn:rwins}
\end{equation}
where $1[n]$ denotes the array filled with the number 1.

\section{Short-time Variational Mode Decomposition}
\label{sec:4}
The original VMD algorithm is a special case of MVMD when the number of time series is equal to one. In the following section, we focus on the general MVMD case that deals with the multiple time series simultaneously.

As an extension of the original VMD and MVMD method, the non-dynamic STVMD redivides the time series into multiple windowed time series. The non-dynamic STVMD hold the same assumption on the central frequencies as that in VMD and MVMD. The main difference between non-dynamic STVMD and VMD or MVMD is introduced by the short-time windows.

The dynamic STVMD, based on the equation given in non-dynamic STVMD, modifies the assumption on the central frequencies. The modification results in the dynamic central frequencies as the time changes.

\subsection{Windowed Multivariate Time Series}
In the multivariate case, where the number of time series is set to $C$, the multivariate signal $\bm{x}[n]\in\mathbb{R}^{C\times T}$ is expressed as $\bm{x}[n] = [x_1[n]; x_2[n];\cdots; x_C[n]]$. As in Equation \ref{eqn:xw}, after applying the window function to $\bm{x}[n]$, the windowed signal is expressed as
\begin{equation}
    \bm{X}[n] =\begin{bmatrix}
                xw_1[n-1] & xw_1[n-2] & \cdots & xw_1[n-T]\\
                xw_2[n-1] & xw_2[n-2] & \cdots & xw_2[n-T]\\
                \vdots    & \vdots    & \ddots & \vdots \\
                xw_C[n-1] & xw_C[n-2] & \cdots & xw_C[n-T]
              \end{bmatrix},
\end{equation}
where $xw_c[n-\tau]=x_c[n]w[n-\tau], c=1,2,...C, \tau=1,2,...,T$, denoting the windowed signal in each time series. 

In the above statements, the discrete format of the Fourier transform is used so that the windowed signals can be presented in a matrix format conveniently. Because $xw[n]$ is obtained by the element-wise product of $x[n]$ and the window function $w[n]$, its relationship still remains when $x$ and $w$ is a continuous signal. The alternative format of the windowed signal is
\begin{equation}
    \bm{X}(t) =\begin{bmatrix}
                xw_{11}(t) & xw_{21}(t) & \cdots & xw_{C1}(t)\\
                xw_{21}(t) & xw_{22}(t) & \cdots & xw_{C2}(t)\\
                \vdots    & \vdots    & \ddots & \vdots \\
                xw_{C1}(t) & xw_{C2}(t) & \cdots & xw_{CT}(t)
              \end{bmatrix},
              \label{eqn:wins}
\end{equation}

The main goal of STVMD is to extract $K$ number of mode functions $\bm{u}_k(t)$ from the input data $\bm{x}[n]$
\begin{equation}
    \bm{X}(t)=\sum_{k=1}^{K}\bm{U}_k(t),
\end{equation}
where
\begin{equation}
    \bm{U}_k(t)=\begin{bmatrix}
                u_{k,11}(t) & u_{k,11}(t) & \cdots & u_{k,11}(t)\\
                u_{k,21}(t) & u_{k,22}(t) & \cdots & u_{k,2T}(t)\\
                \vdots    & \vdots    & \ddots & \vdots \\
                u_{k,C1}(t) & u_{k,C2}(t) & \cdots & u_{k,CT}(t)
              \end{bmatrix}.
\end{equation}
$u_{k,c\tau}(t)$ is the AM-FM signal, and can be expressed as
\begin{align*}
    u_{k,c\tau}(t)&=a_{k,c\tau}(t)cos(\phi_{k,c\tau}(t))\\
    &=\mathcal{R}\{a_{k,c\tau}(t)e^{j\phi_{k,c\tau}(t)}\}
\end{align*}
where $a_{k,c\tau}(t)$ and $\phi_{k,c\tau}(t)$ denote amplitude and phase function corresponding to the $c$-th time series and $\tau$-th time window in $k$-th mode function respectively. The analytic signal of $\bm{u}_k(t)$ is given by
\begin{align*}
    \bm{U}_+^k(t) & = \bm{U}_k(t) + j \mathcal{H}\{\bm{U}_k(t)\}.
\end{align*}

With the input signal $\bm{x}(t)$ or $\bm{x}[n]$ and the output decomposed mode functions $\bm{U}_k(t)$, the goal is to extract the AM-FM signal set $\{\bm{U}_k(t)\}$ from $\bm{x}(t)$ and fulfill the constraints:
\begin{itemize}
    \item the $\bm{x}(t)$ is exactly recovered by summing up the decomposed mode functions;
    \item the sum of bandwidths of each mode function is minimized.
\end{itemize}
The constraints are the same as those in VMD and result in the objective function but in the vector format:
\begin{equation}
f = \sum_{k=1}^K \left\| \partial_t \left[\bm{U}_+^k(t) e^{-j \omega_k t} \right] \right\|_2^2.
\end{equation}

\subsection{Non-dynamic Short-time Variational Mode Decomposition}

VMD has $K$ central frequencies $\omega_k$ for a single time series. In MVMD, multiple time series are analyzed simultaneously. Considering the harmonic mixing of the signals, MVMD has the common central frequencies ${\omega_k}$ in all time series. In the non-dynamic STVMD, both multiple time series and sliding time windows are considered, with common central frequencies $\omega_{k}$. This leads to a convenient representation of the objective function:
\begin{equation}
    f = \sum_{k=1}^K\sum_{c=1}^C\sum_{\tau=1}^T \left\| \partial_t \left[u_+^{k,c\tau}(t) e^{-j \omega_{k} t} \right] \right\|_2^2,
\end{equation}
where $u_+^{k,c\tau}(t)$ is the analytic signal of $u_{k,c\tau}(t)$ in the $k$-th mode function. Now we have the constraint optimization problem for STVMD
\begin{equation}
\min_{\{u_k\}, \{\omega_{k}\}} \left\{ \sum_{k=1}^K\sum_{c=1}^C\sum_{\tau=1}^T \left\| \partial_t \left[u_+^{k,c\tau}(t) e^{-j \omega_{k} t} \right] \right\|_2^2 \right\},
\end{equation}
subject to:
\begin{equation}
\sum_{k=1}^K u_{k,c\tau}(t) = xw_{c\tau}(t),
\end{equation}
where $c=1,2,...,C$, $\tau=1,2,...,T$. The corresponding  augmented Lagrangian function then becomes
\begin{align}
    \mathcal{L}(\{u_{k,c\tau}\}, &\{\omega_{k}\}, \lambda) = \alpha\sum_{k=1}^K\sum_{c=1}^C\sum_{\tau=1}^T \left\| \partial_t \left[u_+^{k,c\tau}(t) e^{-j \omega_{k} t} \right] \right\|_2^2 \nonumber \\
    & + \sum_{c=1}^C\sum_{\tau=1}^T\left\| xw_{c\tau}(t) - \sum_{k=1}^K u_{k,c\tau}(t) \right\|_2^2 \nonumber\\
    & + \sum_{c=1}^C\sum_{\tau=1}^T\left< \lambda_{c\tau}(t), xw_{c\tau}(t) - \sum_{k=1}^K u_{k,c\tau}(t) \right>.
    \label{eqn:objdstvmd}
\end{align}

The above optimization problem is solved using ADMM approach as given in Algorithm \ref{alg:3}. After the optimization process presented in Algorithm \ref{alg:2}, the windowed signal $\bm{x}(t)$ in Equation \ref{eqn:wins} is obtained. The original signal $x(t)$ (or $x[n]$) is reverted with Equation \ref{eqn:rwins}.

\begin{algorithm}[htbp]
\caption{Complete optimization of non-dynamic STVMD}
\begin{algorithmic}
\scriptsize
\STATE Initialize \( \{\hat{u}_{k,c\tau}^1\}, \{\omega_{k}^1\}, \hat{\lambda}_{c\tau}^1, n \leftarrow 0, \epsilon \leftarrow 1e^{-9}\)
\REPEAT
    \STATE \( n \leftarrow n + 1 \)
    \FOR{ $k = 1 : K$ }
    \FOR{ $c = 1 : C$ }
    \FOR{ $\tau = 1 : T$ }
        \STATE \textbf{Update \( \hat{u}_{k,c\tau} \):}
        \STATE
        $\hat{u}_{k,c\tau}^{n+1}(\omega) \leftarrow$
        \begin{equation}
            \frac{\hat{xw}_{c\tau}(\omega) - \sum_{i<k} \hat{u}_{i,c\tau}^{n+1}(\omega) - \sum_{i>k} \hat{u}_{i,c\tau}^n(\omega) + \frac{\hat{\lambda}_{c\tau}^n(\omega)}{2}}{1 + 2\alpha (\omega - \omega^n_k)^2}
            \nonumber
        \end{equation}
    \ENDFOR
    \ENDFOR
    \ENDFOR
    \FOR{ $k = 1 : K$ }
        \STATE \textbf{Update \( \omega_{k} \):}
        \begin{equation}
            \omega_{k}^{n+1} \leftarrow \frac{\sum_{c=1}^C\sum_{\tau=1}^T\int_0^\infty \omega \left| \hat{u}_{k,c\tau}^{n+1}(\omega) \right|^2 d\omega}{\sum_{c=1}^C\sum_{\tau=1}^T\int_0^\infty \left| \hat{u}_{k,c\tau}^{n+1}(\omega) \right|^2 d\omega}.
            \nonumber
        \end{equation}
    \ENDFOR
    \FOR{ $c = 1 : C$ }
    \FOR{ $\tau = 1 : T$ }
        \STATE \textbf{Update \( \hat{\lambda}_{c\tau} \):}
        \begin{equation}
        \hat{\lambda}_{c\tau}^{n+1}(\omega) \leftarrow \hat{\lambda}_{c\tau}^n(\omega) + \tau \left( \hat{xw}_{c\tau}(\omega) - \sum_{k} \hat{u}_{k,c\tau}^{n+1}(\omega) \right)
        \nonumber
        \end{equation}
    \ENDFOR
    \ENDFOR
\UNTIL{convergence: \( \max \frac{\|\hat{u}_{k,c}^{n+1} - \hat{u}_{k,c}^n\|_2^2}{\|\hat{u}_{k,c}^n\|_2^2} < \epsilon, c=1,2,\cdots,C\).}
\end{algorithmic}
\label{alg:3}
\end{algorithm}

\subsection{Dynamic Short-time Variational Mode Decomposition}

In the non-dynamic STVMD, the central frequencies $\omega$ are the same for all time series and short time windows. In the multivariate signals, the multiple time series are acquired simultaneously. This helps to support the assumption in MVMD that multiple time series share the same central frequencies. However, time series are dynamic as time changes, indicating that the central frequencies are not always centered at fixed values. To adapt to the dynamic property of time series, the dynamic STVMD is proposed. In the dynamic STVMD, the central frequencies change as the time window changes while all the time series share the same central frequencies within a time window. Compared to the central frequencies $\bm{\omega}\in \mathbb{R}^{K\times1}$ used in VMD, MVMD and non-dynamic STVMD, dynamic STVMD has the central frequencies $\bm{\Omega}\in \mathbb{R}^{K\times T}$. The complete optimization of the dynamic STVMD is given in Algorithm \ref{alg:4}.

The objective function of dynamic STVMD in Lagrange form is given by
\begin{align}
    \mathcal{L}(\{u_{k,c\tau}\}, &\{\omega_{k,\tau}\}, \lambda) = \alpha\sum_{k=1}^K\sum_{c=1}^C\sum_{\tau=1}^T \left\| \partial_t \left[u_+^{k,c\tau}(t) e^{-j \omega_{k,\tau} t} \right] \right\|_2^2 \nonumber \\
    & + \sum_{c=1}^C\sum_{\tau=1}^T\left\| xw_{c\tau}(t) - \sum_{k=1}^K u_{k,c\tau}(t) \right\|_2^2 \nonumber\\
    & + \sum_{c=1}^C\sum_{\tau=1}^T\left< \lambda_{c\tau}(t), xw_{c\tau}(t) - \sum_{k=1}^K u_{k,c\tau}(t) \right>.
\end{align}
The dynamic central frequencies $\omega_{k, \tau}$ is the element of $\bm{\Omega}$, which leads to the changes of solving process in Equation \ref{eqn:uk} and Equation \ref{eqn:wk}.The updating process of $\hat{u}_{k,c\tau}$ is new presented with
\begin{equation}
    \hat{u}_{k,c\tau}^{n+1}(\omega)=\frac{\hat{xw}-\sum_{i!=k}\hat{u}_{k,c\tau}(\omega)+\frac{\hat{\lambda(\omega)}}{2}}{1+2\alpha(\omega-\omega_{k,\tau})^2},
\end{equation}
and the updating process of central frequencies $\omega_{k,\tau}$ is given by
\begin{equation}
    \omega_{k,\tau}^{n+1} \leftarrow \frac{\sum_{c=1}^C\int_0^\infty \omega \left| \hat{u}_{k,\tau}^{n+1}(\omega) \right|^2 d\omega}{\sum_{c=1}^C\int_0^\infty \left| \hat{u}_{k,\tau}^{n+1}(\omega) \right|^2 d\omega}.
\end{equation} 

\begin{algorithm}[htbp]
\caption{Complete optimization of dynamic STVMD}
\begin{algorithmic}
\scriptsize
\STATE Initialize \( \{\hat{u}_{k,c\tau}^1\}, \{\omega_{k,\tau}^1\}, \hat{\lambda}_{c\tau}^1, n \leftarrow 0, \epsilon \leftarrow 1e^{-9}\)
\REPEAT
    \STATE \( n \leftarrow n + 1 \)
    \FOR{ $k = 1 : K$ }
    \FOR{ $c = 1 : C$ }
    \FOR{ $\tau = 1 : T$ }
        \STATE \textbf{Update \( \hat{u}_{k,c\tau} \):}
        \STATE
        $\hat{u}_{k,c\tau}^{n+1}(\omega) \leftarrow$
        \begin{equation}
            \frac{\hat{xw}_{c\tau}(\omega) - \sum_{i<k} \hat{u}_{i,c\tau}^{n+1}(\omega) - \sum_{i>k} \hat{u}_{i,c\tau}^n(\omega) + \frac{\hat{\lambda}_{c\tau}^n(\omega)}{2}}{1 + 2\alpha (\omega - \omega^n_{k,\tau})^2}
            \nonumber
        \end{equation}
    \ENDFOR
    \ENDFOR
    \ENDFOR
    \FOR{ $k = 1 : K$ }
    \FOR{ $\tau = 1 : T$ }
        \STATE \textbf{Update \( \omega_{k,\tau} \):}
        \begin{equation}
            \omega_{k,\tau}^{n+1} \leftarrow \frac{\sum_{c=1}^C\int_0^\infty \omega \left| \hat{u}_{k,c\tau}^{n+1}(\omega) \right|^2 d\omega}{\sum_{c=1}^C\int_0^\infty \left| \hat{u}_{k,c\tau}^{n+1}(\omega) \right|^2 d\omega}.
            \nonumber
        \end{equation}
    \ENDFOR
    \ENDFOR
    \FOR{ $c = 1 : C$ }
    \FOR{ $\tau = 1 : T$ }
        \STATE \textbf{Update \( \hat{\lambda}_{c\tau} \):}
        \begin{equation}
        \hat{\lambda}_{c\tau}^{n+1}(\omega) \leftarrow \hat{\lambda}_{c\tau}^n(\omega) + \tau \left( \hat{xw}_{c\tau}(\omega) - \sum_{k} \hat{u}_{k,c\tau}^{n+1}(\omega) \right)
        \nonumber
        \end{equation}
    \ENDFOR
    \ENDFOR
\UNTIL{convergence: \( \max \frac{\|\hat{u}_{k,c}^{n+1} - \hat{u}_{k,c}^n\|_2^2}{\|\hat{u}_{k,c}^n\|_2^2} < \epsilon, c=1,2,\cdots,C\).}
\end{algorithmic}
\label{alg:4}
\end{algorithm}

The main difference between non-dynamic STVMD and dynamic STVMD is in the central frequencies. The non-dynamic STVMD is a special case of dynamic STVMD where the central frequencies are kept as time changes.

\subsection{Online Short-time Variational Mode Decompostion}
Offline process refers to the processing of data that has already been collected and stored. Online processing refers to real-time processing of incoming data signals, which involves analysing and manipulating the data as it is collected, without the need to store the entire data set before starting the processing task.

In VMD and MVMD, the entire time series are fed into the Fourier transform for the decomposition of signal in the Fourier domain. This operation indicates that VMD and MVMD are offline signal processing algorithms, and suffer from heavy computation load in the online process because of the computation of the whole dataset.

In the dynamic STVMD, the central frequencies are independent for two non-adjacent time windows. In the online process, new signals are added to the original signals. The newly added signals can influence the central frequencies of all signals, so it is necessary to update the central frequencies. In dynamic STVMD, due to the use of short time windows, the influence of new signals is limited to the latest $N$ sampling points, where $N$ is the length of the time window. The pseudo computation process is given in Algorithm \ref{alg:5}. In the offline process, before applying STVMD, the reflection padding on both left and right sides is used to apply the window function on the signals. But in the online process, the right-side reflection padding is used because the cached signals are used on the left side.

\begin{algorithm}[htbp]
\caption{Psedo Online dynamic STVMD process}
\begin{algorithmic}
\STATE Given the offline cached signal $\bm{x}_{old}[n]\in\mathbb{R}^{C\times{(N-1)}}$ and the online signal $\bm{x}_{new}[n]\in\mathbb{R}^{C\times{1}}$;

\STATE Append $\bm{x}_{new}$ to $\bm{x}_{old}$ and obtain $\bm{x}[n]\in\mathbb{R}^{C\times N}$;

\STATE Apply reflection padding to $\bm{x}[n]$ on the right side;

\STATE Apply dynamic STVMD in Algorithm \ref{alg:4}.

\end{algorithmic}
\label{alg:5}
\end{algorithm}

\section{Experimental Results}
\label{sec:5}
In this section, experiments and simulations are used to evaluate the performance of the proposed STVMD method, and the detailed results are given.

The experiments include two points: (1) Comparison between VMD (or MVMD) and non-dynamic STVMD, to analyze the influence induced by the short time windows; (2) New property of decomposed signals in the dynamic STVMD compared to non-dynamic STVMD. The default parameter setting for both VMD (or MVMD) and STVMD is given in Table \ref{tab:1}.

\begin{table}[htbp]
    \centering
    \scriptsize
    \caption{Default Parameter Setting for VMD, MVMD and STVMD}
    \begin{tabular}{c|c|c|c|c}
       \toprule
       Parameters  & $\alpha$ & $\omega_{init}$ & $\epsilon$ & window type \\
       \midrule
       VMD/MVMD/STVMD  & 50 & uniform distribution & $10^{-9}$ & Hamming \\
       \bottomrule
    \end{tabular}
    \label{tab:1}
\end{table}

In the initialization of $\omega$ with uniform distribution, the uniform distribution ranges from 0 to 1 in an arithmetic sequence, $i.e.$ $0, \frac{1}{K}, \frac{2}{K}, \cdots, \frac{K-1}{K}$, where $K$ is the number of mode functions. During the updating process of VMD or STVMD, the central frequency at 0$Hz$ is not updated in the loop and remains zero. The mode function at 0$Hz$ is a residual mode, and is not displayed in the result analysis if not mentioned specifically. 

\subsection{Non-dynamic STVMD}

In VMD, the mode decomposition is applied to the Fourier transform of the entire time series. Compared to VMD, non-dynamic STVMD modifies the range of time series with the short time windows. The experiments in this subsection focus on the influence induced by the short time window, $i.e.$, the influence of the length of the time window and the mode alignment in the multivariate signals.

\subsubsection{Influence of Length of Time Window}
\begin{figure}[htbp]
    \centering
    \includegraphics[width=0.485\textwidth]{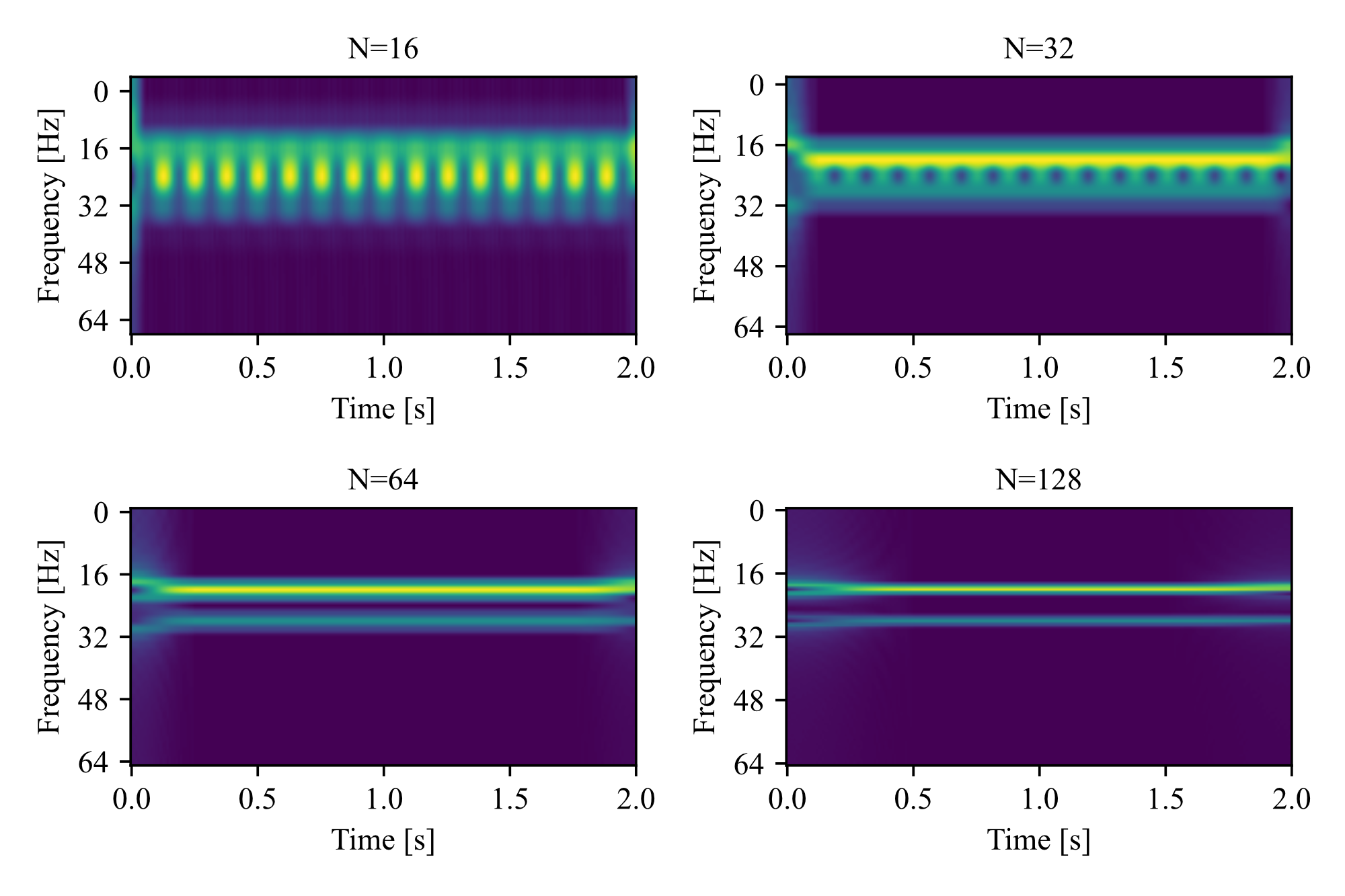}
    \caption{The time-frequency spectrum of the input signal of non-dynamic STVMD when the length of Hamming window is 16, 32, 64, 128, respectively. }
    \label{fig:1.1a}
\end{figure}

\begin{figure}[htbp]
    \centering
    \includegraphics[width=0.485\textwidth]{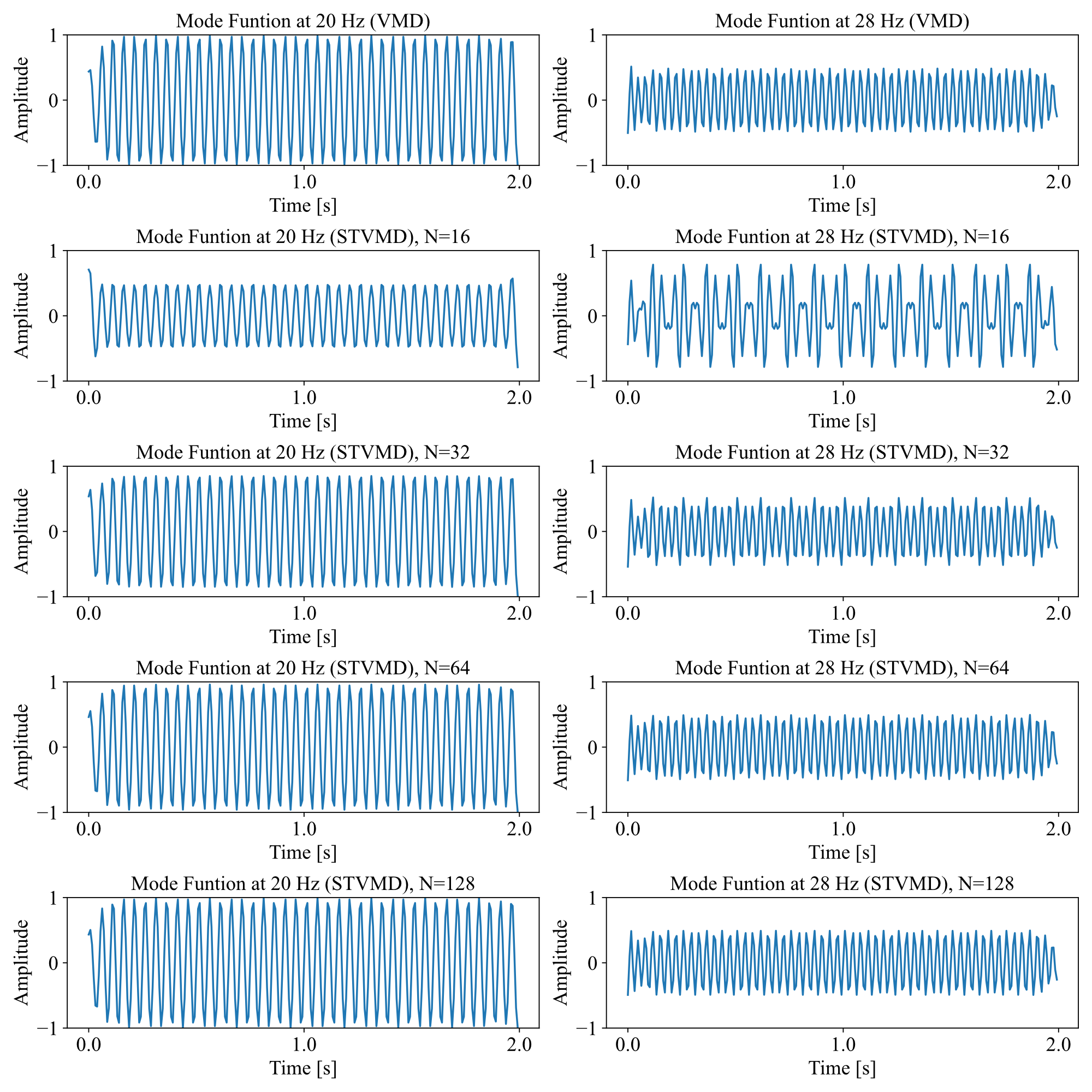}
    \caption{Decomposed mode functions with VMD and non-dynamic STVMD. The input is a mixture of 20$Hz$ and 28$Hz$ sinusoid functions. The number of mode functions is 3, and the mode functions whose central frequencies are not zeros are illustrated in the figure. The $N$ refers to the length of the hamming window in STVMD.}
    \label{fig:1.1b}
\end{figure}

In Figure \ref{fig:1.1a}, we illustrate the time-frequency spectrum of the input signals for STVMD, with the time window length $N=16, 32, 64, 128$. 
In Figure \ref{fig:1.1b}, we illustrate the influence of the length of the window function on non-dynamic STVMD. The input data is a single time series with a sampling rate of 128$Hz$. The individual components of the input signal are a mixture of a 20$Hz$ sinusoid with amplitude 1 and 28$Hz$ sinusoid with amplitude 0.5, $e.g.$,
\begin{equation}
    x(t)=sin(2\pi*20*t)+0.5*sin(2\pi*28*t).
    \label{eqn:ssinu}
\end{equation} 
In Figure \ref{fig:1.1b}, the mode functions with non-zero central frequencies are given for both VMD and non-dynamic STVMD. In the non-dynamic STVMD, the length of window function $N$ changes from 16 to 128, and meanwhile the signal waveform approaches the one decomposed with VMD.


\subsubsection{Mode Alignment Property}
The alignment of multivariate data or multiple time series, termed mode alignment, involves synchronizing common or joint oscillations, which are characterized by similar frequency content, across multiple time series within a single mode. In MVMD, because all the time series share the same central frequencies, MVMD has the property of mode alignment. Non-dynamic inherits the mode alignment of MVMD in the decomposition of multivariate time series. 

\begin{figure}[htbp]
    \centering
    \includegraphics[width=0.485\textwidth]{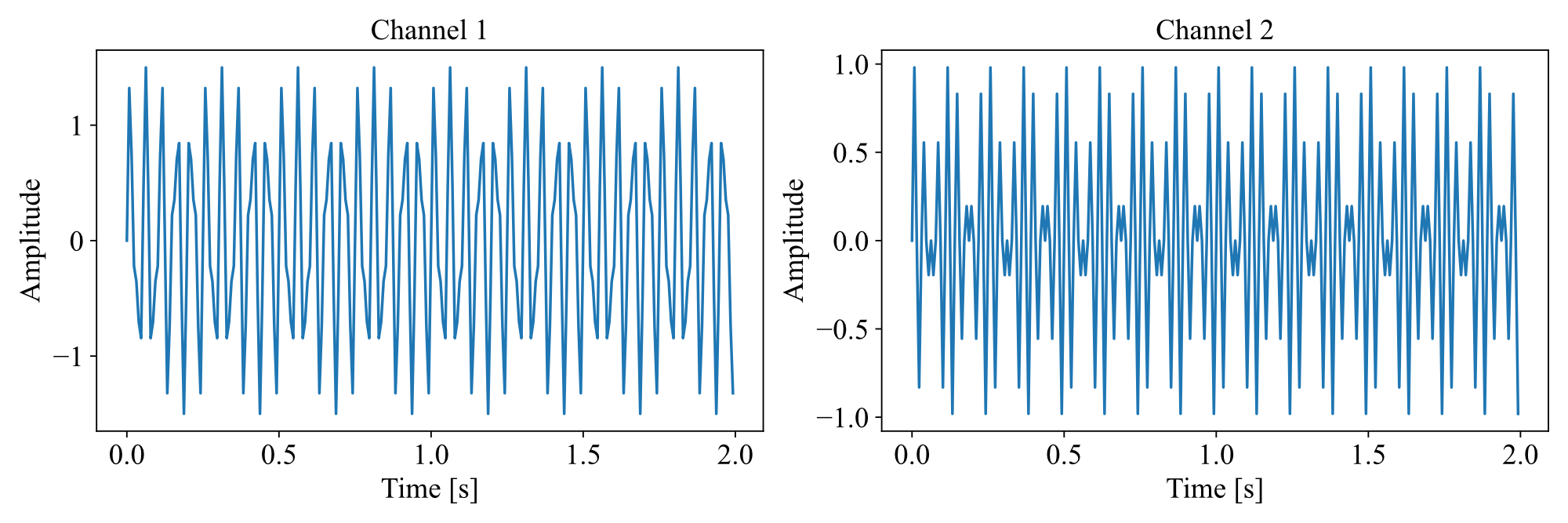}
    \caption{The input two time series with sampling rate 128$Hz$, which is used to validate the mode alignment of non-dynamic STVMD.}
    \label{fig:1.2a}
\end{figure}

\begin{figure*}[htbp]
    \centering
    \includegraphics[width=\textwidth]{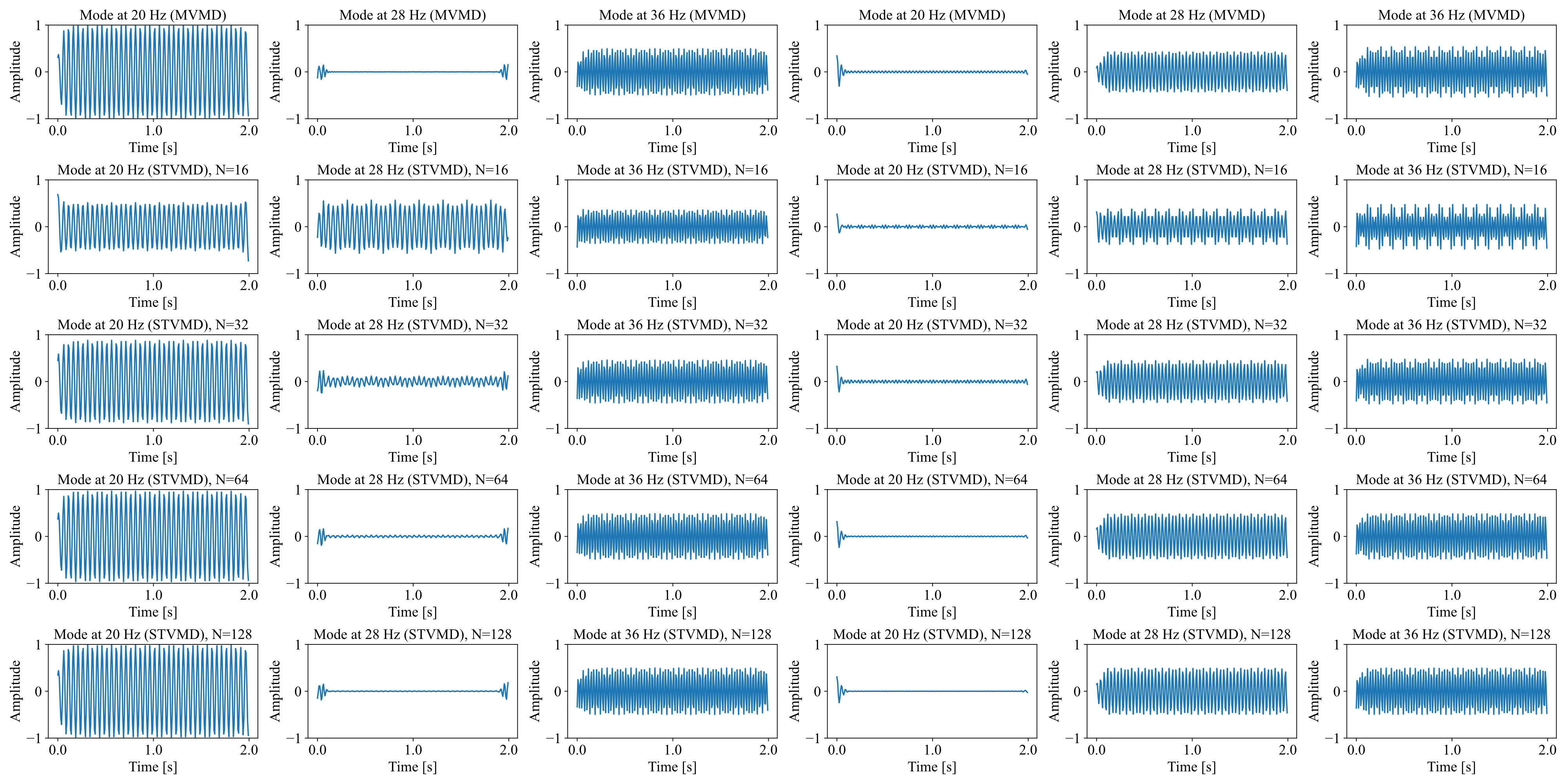}
    \caption{The mode functions decomposed with MVMD and non-dynamic STVMD. $N$ is the length of time window used in the STVMD. Mode functions at frequencies 20$Hz$, 28$Hz$ and 36$Hz$ are given in this figure. The left three columns are decomposed from Channel 1 and the right three columns are decomposed from Channel 2.}
    \label{fig:1.2c}
\end{figure*}

Figure \ref{fig:1.2a} is the input two time series sampled at 128$Hz$ to validate the mode alignment of non-dynamic STVMD. The two time series share the common 36 $Hz$ sinusoid with an amplitude of 0.5. The two time series are given by the following equation, 
\begin{equation}
    \bm{x}(t) = \begin{bmatrix}
    \sin(2\pi \cdot 20 \cdot t) + 0.5 \sin(2\pi \cdot 36 \cdot t) \\
    \sin(2\pi \cdot 28 \cdot t) + 0.5 \sin(2\pi \cdot 36 \cdot t)
    \end{bmatrix}.
\end{equation}
In Figure \ref{fig:1.2c}, the time-frequency spectrums of two time series are illustrated. The spectrum is calculated by STFT with length of time window $N$=16, 32, 64, 128. Figure \ref{fig:1.2c} compares the mode functions decomposed by MVMD and non-dynamic STVMD.
\begin{figure}[htbp]
    \centering
    \includegraphics[width=0.485\textwidth]{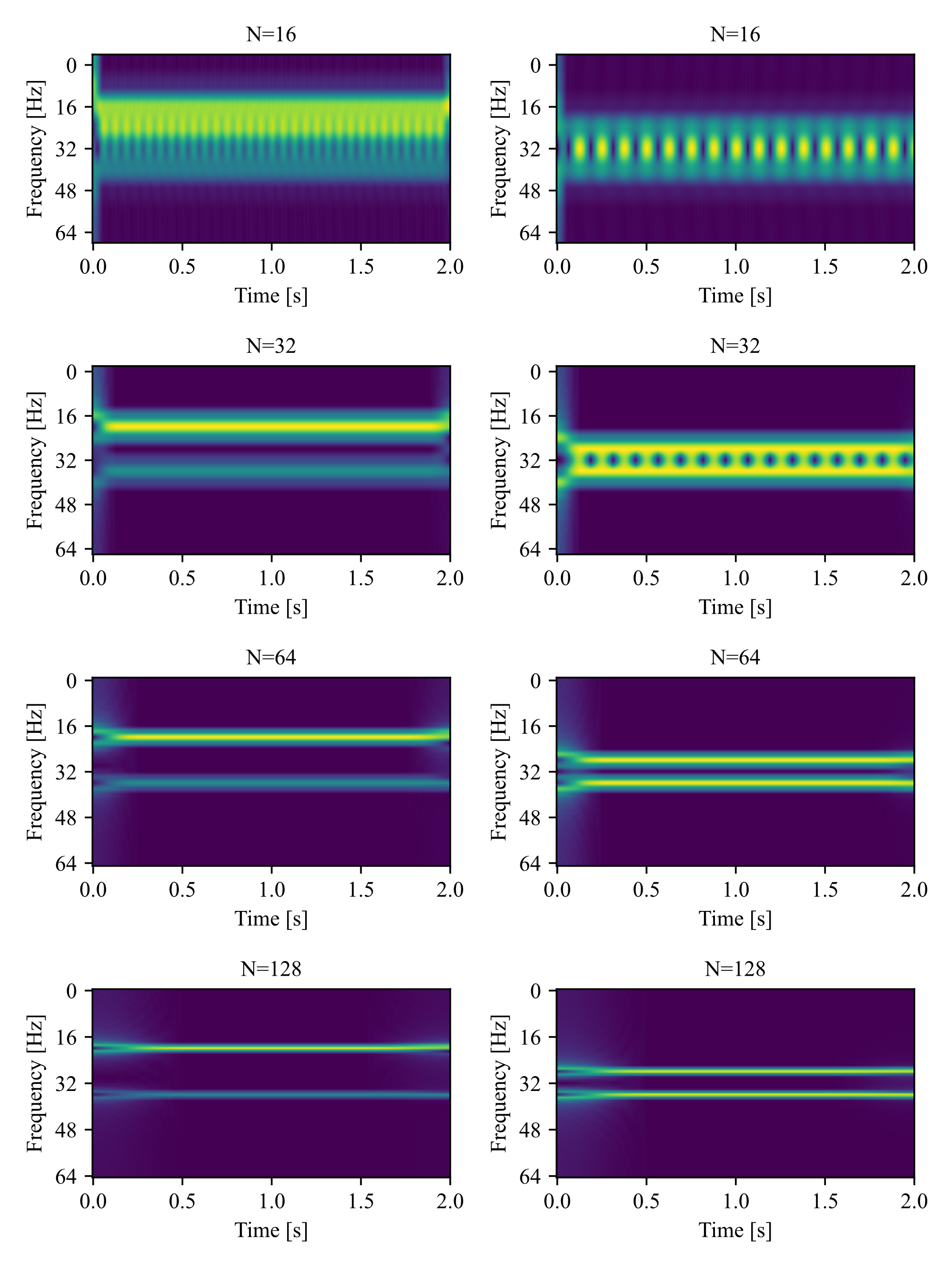}
    \caption{The time-frequency spectrum of the two time series in Figure \ref{fig:1.2a}, Channel 1: left column, Channel 2: right column. The spectrum is calculated with STFT and $N$ is the length of time windows in  STFT.}
    \label{fig:1.2b}
\end{figure}

In the above comparison between VMD and non-dynamic STVMD, we can observe that:

\noindent (1) The non-dynamic STVMD inherits the mode decomposition and mode alignment of VMD and MVMD;

\noindent (2) The decomposition performance of non-dynamic STVMD is influenced by the length of time windows on STFT. The reason is that the time window influences the resolution of the time-frequency spectrum in STFT.

\subsection{Non-dynamic and Dynamic STVMD}
In the non-dynamic STVMD, the central frequencies $\omega_k$ remain stationary as time changes. However, the real-world signals are non-stationary, which means that frequencies of mode functions may vary as time changes. The dynamic STVMD is proposed under the assumption on the varying central frequencies. To evaluate the relationship between non-dynamic and dynamic STVMD, the performance of dynamic STVMD is validated in both the stationary and non-stationary simulated signals.

\subsubsection{Dynamic STVMD in Stationary Signals}
\begin{figure}[htbp]
    \centering
    \includegraphics[width=0.485\textwidth]{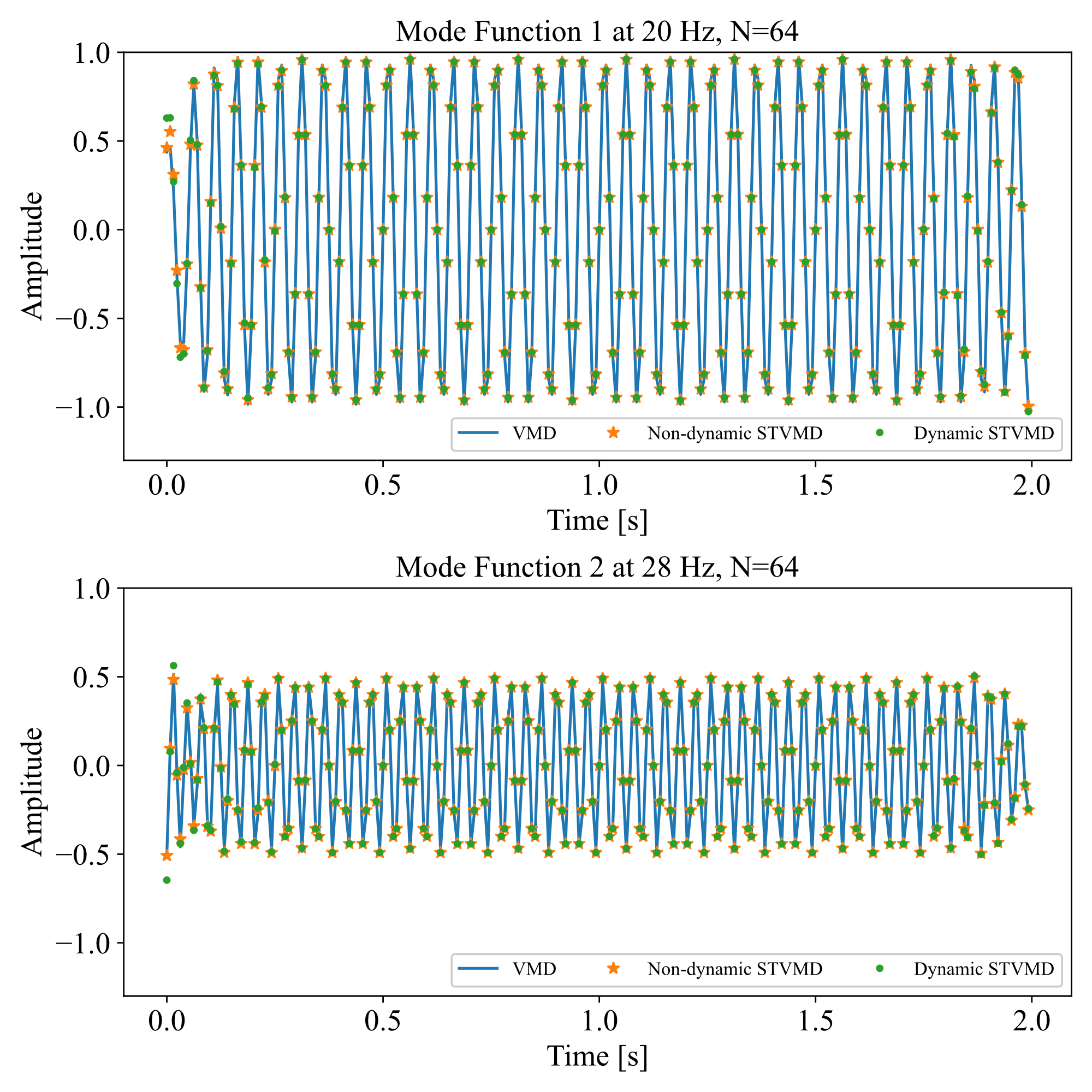}
    \caption{The comparison of mode functions between VMD and STVMD in the stationary signals. The length of time window is set to 64 in both non-dynamic STVMD and dynamic STVMD.}
    \label{fig:2.1a}
\end{figure}

\begin{figure}[htbp]
    \centering
    \includegraphics[width=0.485\textwidth]{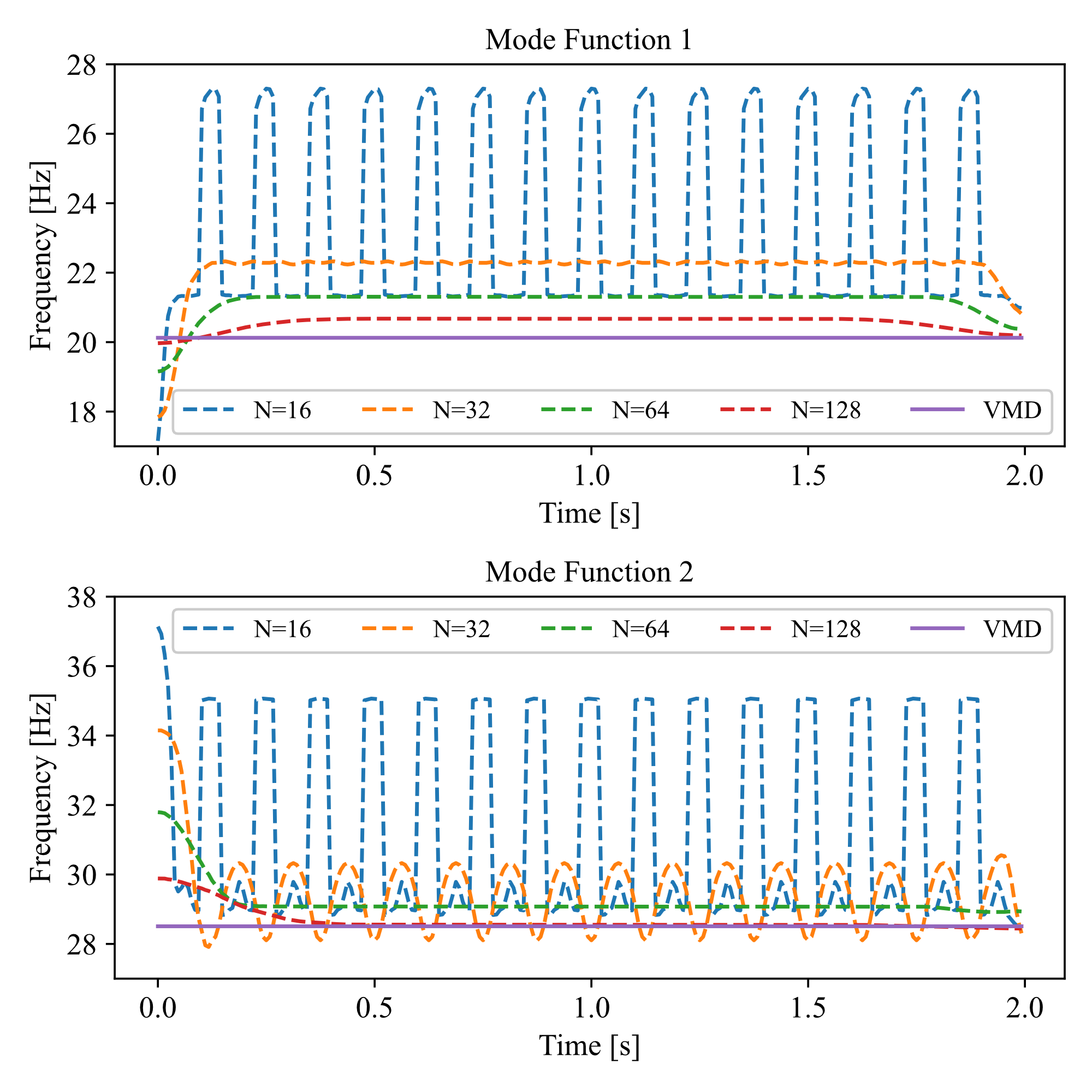}
    \caption{The comparison of central frequencies of non-dynamic STVMD in the stationary signals. The result shows that as the increase of length of time windows, the central frequency of STVMD is approaching to that of VMD.}
    \label{fig:2.1b}
\end{figure}

In the comparison of non-dynamic and dynamic STVMD, the signal in Equation \ref{eqn:ssinu} is used to analyze how the central frequencies are influenced by the length of time windows. Figure \ref{fig:2.1a} compares the results obtained between non-dynamic STVMD and dynamic STVMD for the two decomposed mode functions. Figure \ref{fig:2.1b} gives how the central frequencies change in the decomposition of stationary signals with dynamic STVMD.

\subsubsection{Dynamic STVMD in Non-stationary Signals}
In the non-stationary signal, the frequencies of the signals are time-varying, leading to the necessity of frequency tracking in signal processing. In VMD or MVMD, the central frequencies converge to constant values, which cannot track the dynamic change of the central frequencies of non-stationary signals. In this subsection, three single non-stationary time series are simulated, which are given by

\noindent (1) Simulated signal 1
\begin{align}
     \omega(t) = &\text{{seq}}[\lfloor t \rfloor] + 13 \nonumber\\
     x(t) = &\sin\left(2\pi \omega(t)t\right) + 0.5 \sin\left(2\pi 2\omega(t)t\right) + 0.2 \eta(t);
     \label{eqn:sim1}
\end{align}
where seq is an array $[2, 5, 0, 6, 3, 1, 4, 7]$ and $\lfloor t \rfloor$ denotes the maximum integer smaller than $t$.

\noindent (2) Simulated signal 2
\begin{align}
     x(t) = &\sin\left(2\pi \left(20t + 10\right) t\right) \nonumber\\
     &+ 0.5 \sin\left(2\pi \left(20t + 20\right) t\right) + 0.2 \eta(t);
\end{align}

\noindent (3) Simulated signal 3
\begin{align}
     x(t) = &\sin\left(2\pi \left(2 \sin(2\pi \cdot 0.25 \cdot t) + 10\right) t\right) \nonumber\\
     &+ 0.5 \sin\left(2\pi \left(1.5 \cos(2\pi \cdot 0.25 \cdot t) + 40\right) t\right) \nonumber\\
     &+ 0.2 \eta(t).
\end{align}
In the above equations, $\eta(t)$ refers to the random white noises.

\begin{figure*}[htbp]
    \centering
    \subfigure[Simulated signal 1]{
    \includegraphics[width=\textwidth]{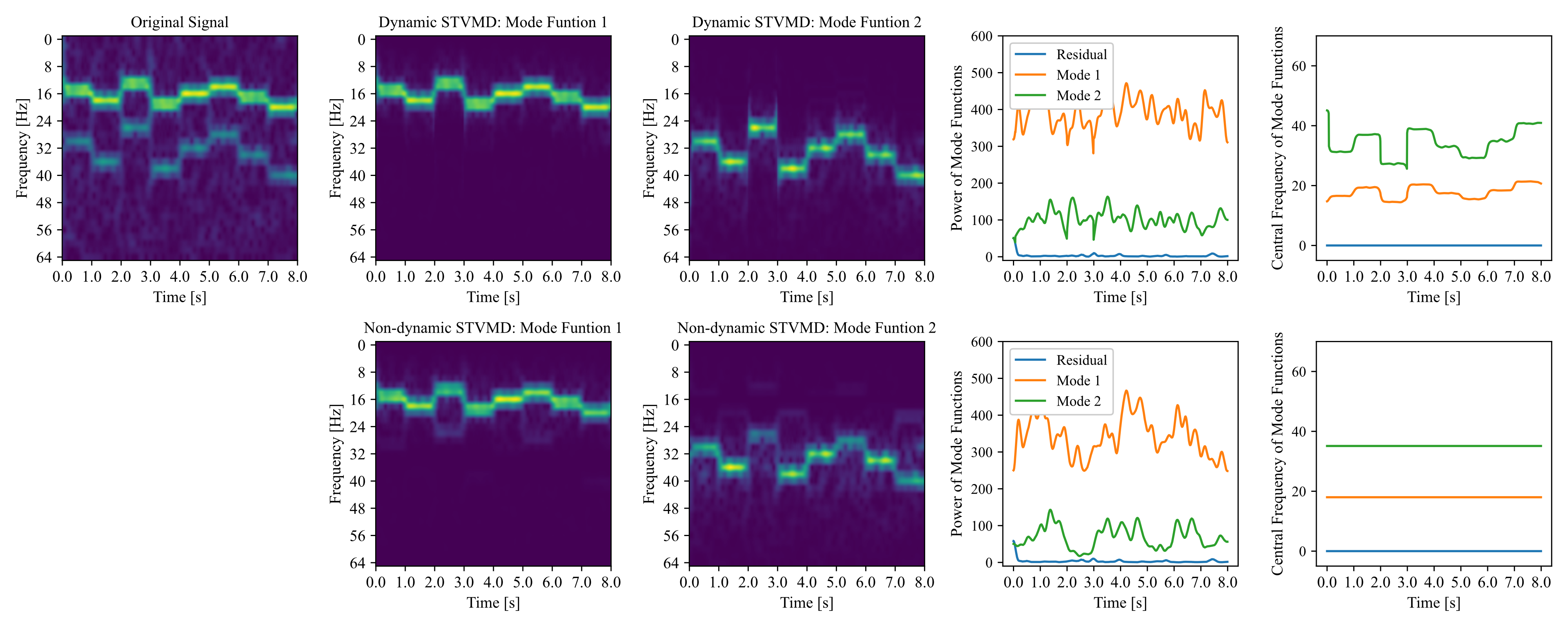}
    \label{fig:2.2a}
    }
    \subfigure[Simulated signal 2]{
    \includegraphics[width=\textwidth]{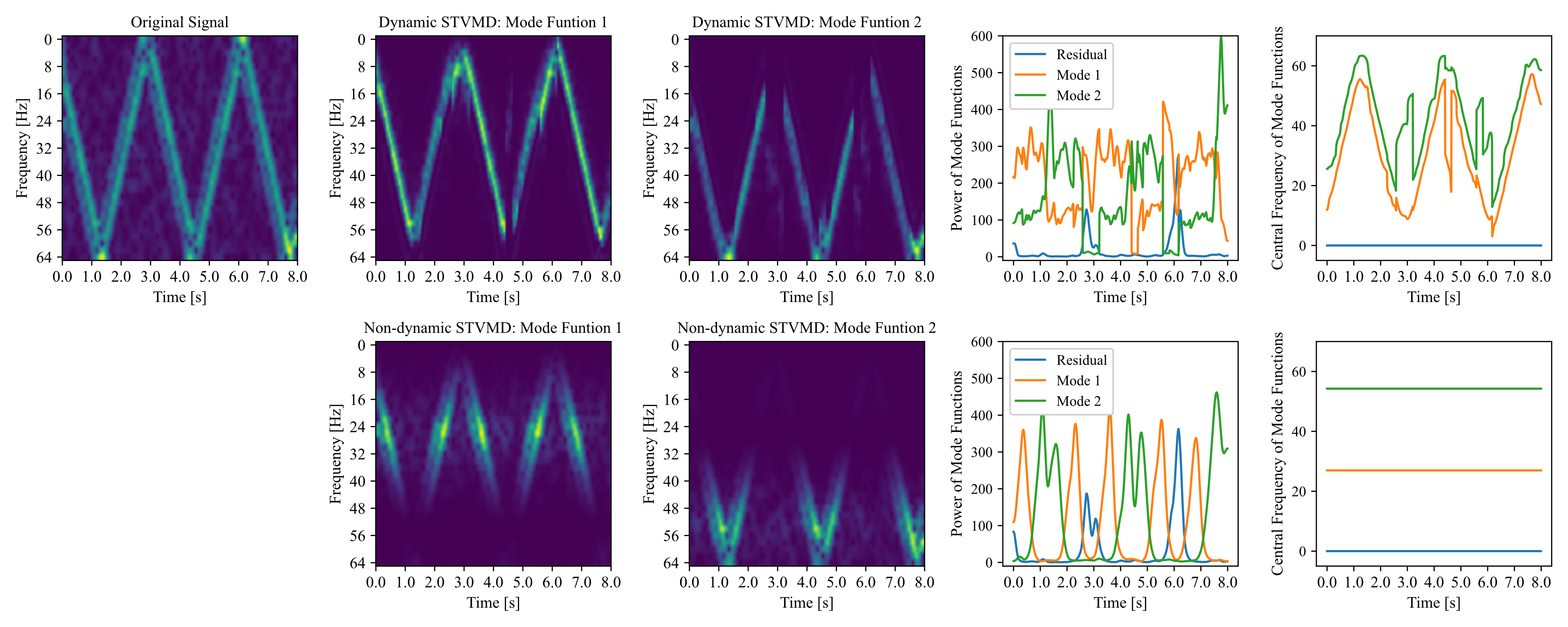}
    \label{fig:2.2b}
    }
    \subfigure[Simulated signal 3]{
    \includegraphics[width=\textwidth]{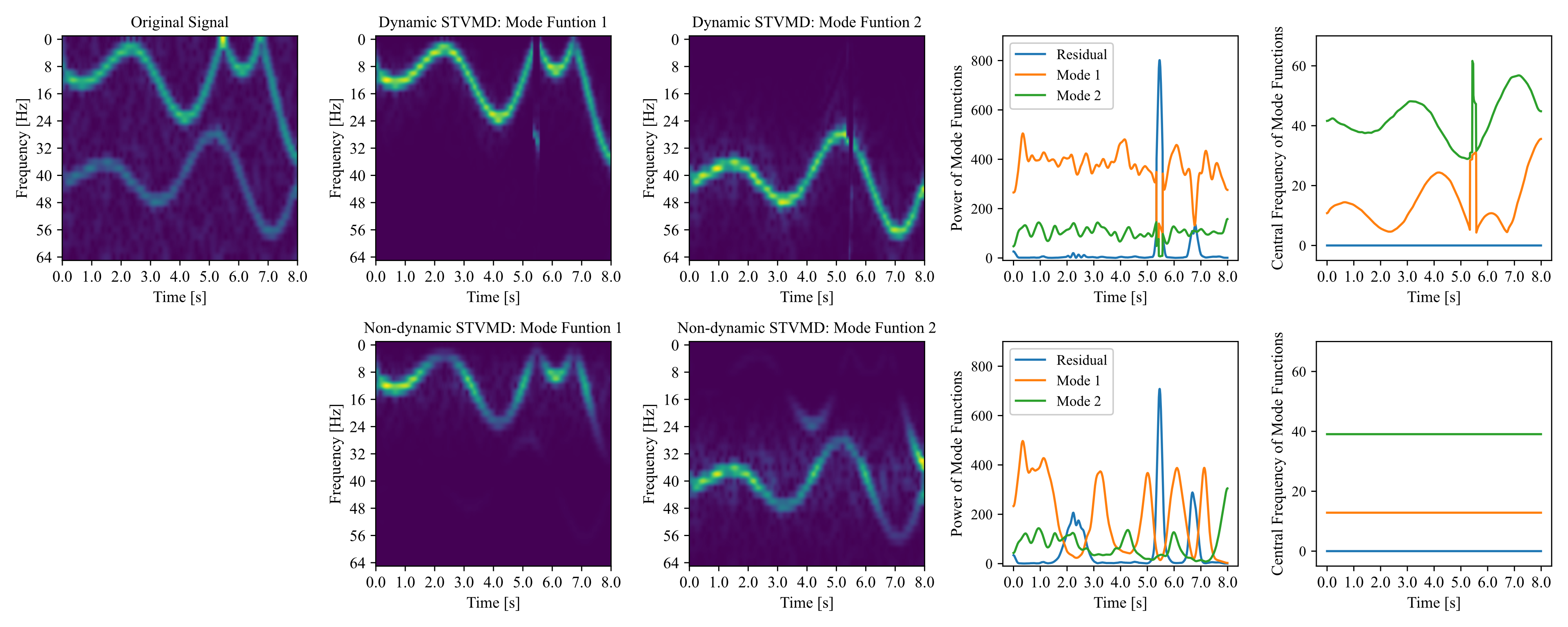}
    \label{fig:2.2c}
    }
    \caption{The decomposition result comparison between non-dynamic STVMD and dynamic STVMD in the simulated signals 1, 2 and 3. In each subfigure, the first column is the time-frequency spectrum of the input signal. The second and third columns are the time-frequency spectrums of the two mode functions, respectively. The fourth column is the power spectrum of the mode functions and the residual at 0$Hz$, which is given by $\left\| \partial_t \left[u_+^{k,c\tau}(t) e^{-j \omega_{k} t} \right] \right\|_2^2$ in Equation \ref{eqn:objdstvmd}. The fifth column shows the central frequencies of the mode functions and the residual. Instead of the constant central frequencies in the non-dynamic STVMD, the central frequencies are time-varying functions in the dynamic STVMD. The frequency changes in the simulated signals are well tracked by the time-varying functions of dynamic STVMD.}
    \label{fig:2.2}
\end{figure*}

\begin{figure*}[htbp]
    \centering
    \subfigure[Simulated signal 1]{
    \includegraphics[width=0.31\textwidth]{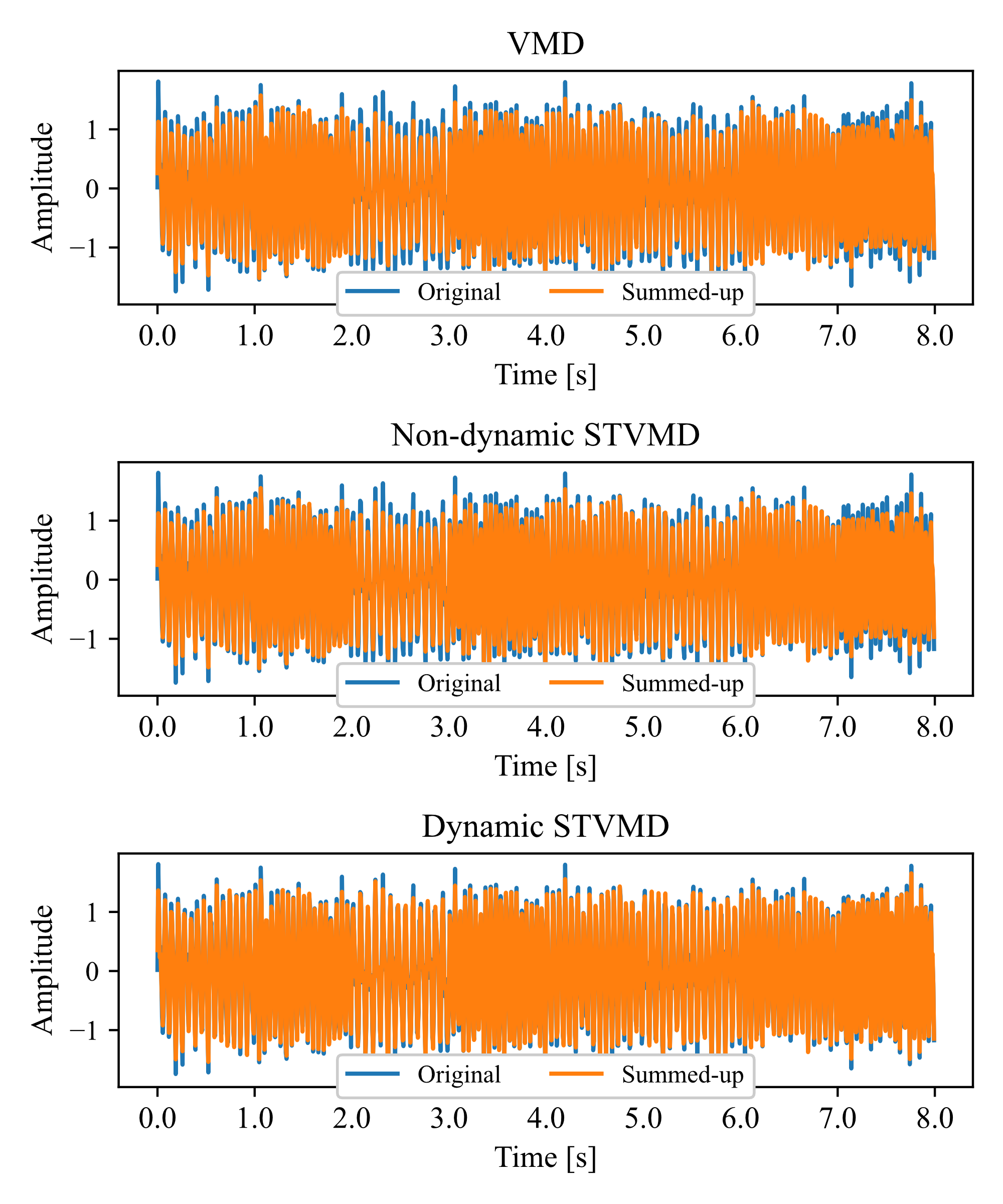}
    \label{fig:2.3a}
    }
    \subfigure[Simulated signal 2]{
    \includegraphics[width=0.31\textwidth]{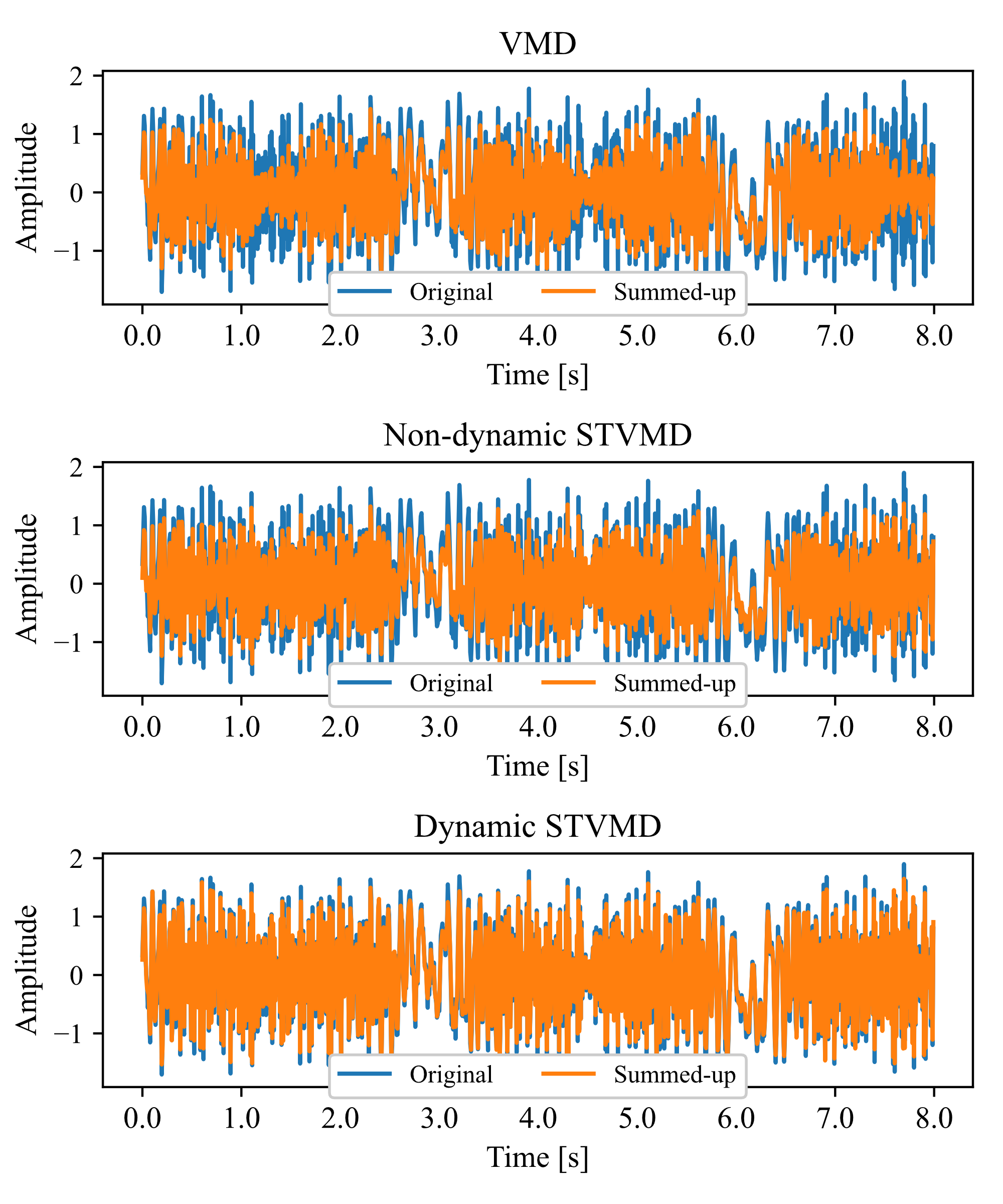}
    \label{fig:2.3b}
    }
    \subfigure[Simulated signal 3]{
    \includegraphics[width=0.31\textwidth]{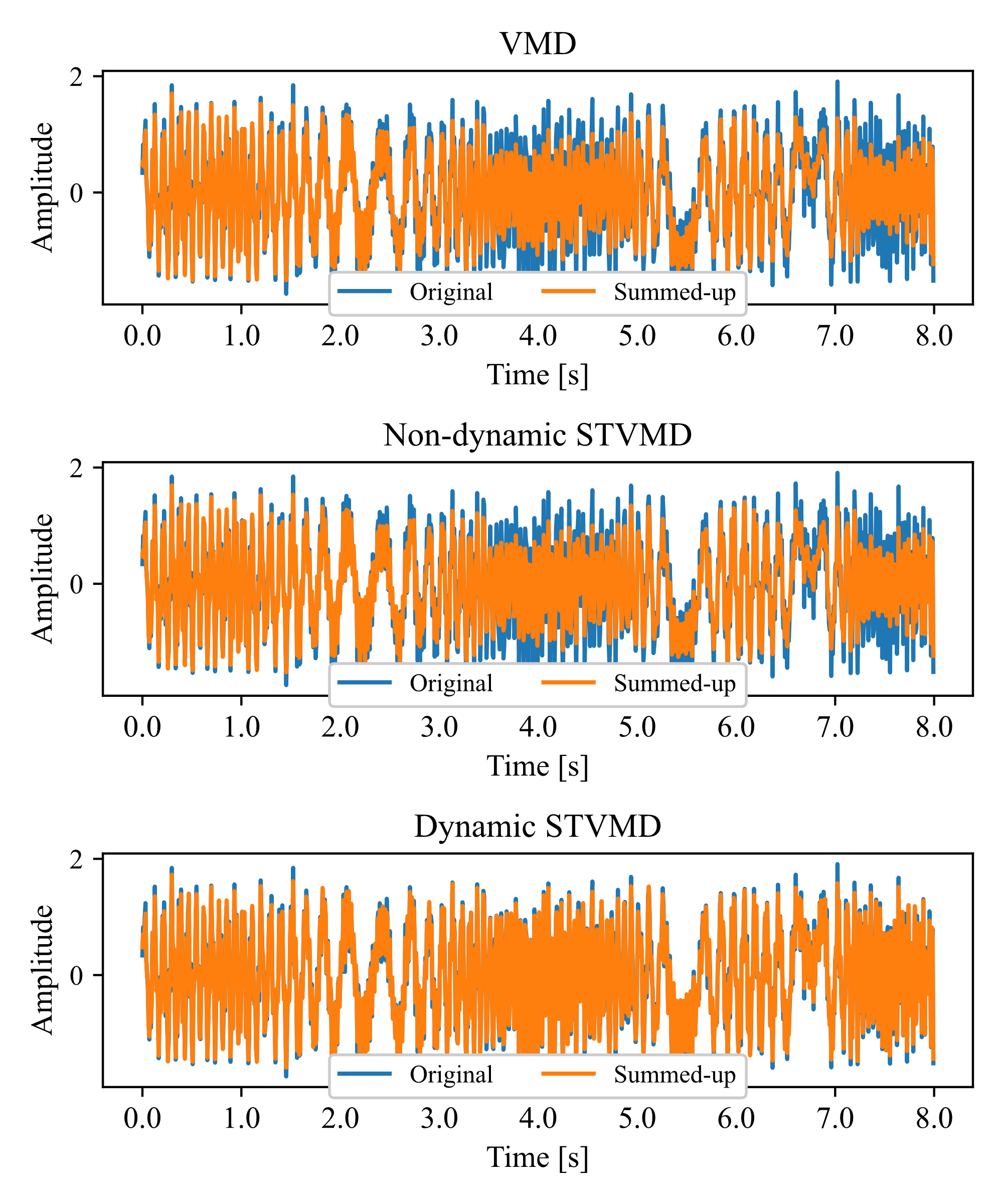}
    \label{fig:2.3c}
    }
    \caption{The comparison between the original input signal and the recovered signal in the simulated signals 1, 2, and 3. The recovered signals are obtained by adding all the mode functions (including the residual). In the decomposition process, VMD\cite{dragomiretskiy_variational_2014}, non-dynamic STVMD and dynamic STVMD share the same parameter setting. In non-stationary signals, the VMD and non-dynamic STVMD with fixed central frequencies cannot fully recover the input signals.}
    \label{fig:2.3}
\end{figure*}

The results of the above three simulated signals, decomposed with non-dynamic STVMD and dynamic STVMD, are illustrated in Figure \ref{fig:2.2a}, Figure \ref{fig:2.2b} and Figure \ref{fig:2.2c}, respectively. The comparison between the original input signals and the recovered signals is given in Figure \ref{fig:2.3}. In Table \ref{tab:2}, the root mean square errors between original signal and recovered signal are given for three simulated signals, respectively.

\begin{table}[htbp]
    \centering
    \caption{Root Mean Square Error between Original and Recovered Signal}
    \begin{tabular}{c|c|c|c|c}
        \toprule
        Simulated Signal & 1 & 2 & 3 & Average\\
        \midrule
        VMD\cite{dragomiretskiy_variational_2014} & 0.1239 & 0.2982 & 0.2227 & 0.2149\\
        Non-dynamic STVMD & 0.1220 & 0.2691 & 0.2236 & 0.2049\\
        Dynamic STVMD & \textbf{0.0950} & \textbf{0.1058} & \textbf{0.0886} & \textbf{0.0965}\\
        \bottomrule
    \end{tabular}
    
    \label{tab:2}
\end{table}

\subsubsection{Mode Alignment Property}
In dynamic STVMD, the mode alignment is inherited from the non-dynamic STVMD. In this experiment, two time series are used to simulate the property of mode alignment in dynamic STVMD. The two time series are given by
\begin{equation}
    \begin{bmatrix}
        x_1(t)\\
        x_2(t)
    \end{bmatrix}
    =
    \begin{bmatrix}
        \sin\left(2\pi \omega(t)t\right) + 0.5 \sin\left(2\pi 2\omega(t)t\right) + 0.2\eta(t)\\
        \sin\left(2\pi \omega(t)t\right) + 0.5 \sin\left(2\pi 3\omega(t)t\right) + 0.2\eta(t)
    \end{bmatrix},
\end{equation}
where $\omega(t)$ is the same as that in Equation \ref{eqn:sim1}.
\begin{figure*}[htbp]
    \centering
    \subfigure[The time-frequency spectrum of the original signals and mode functions in the validation of mode alignment of dynamic STVMD.]
    {
    \includegraphics[width=\textwidth]{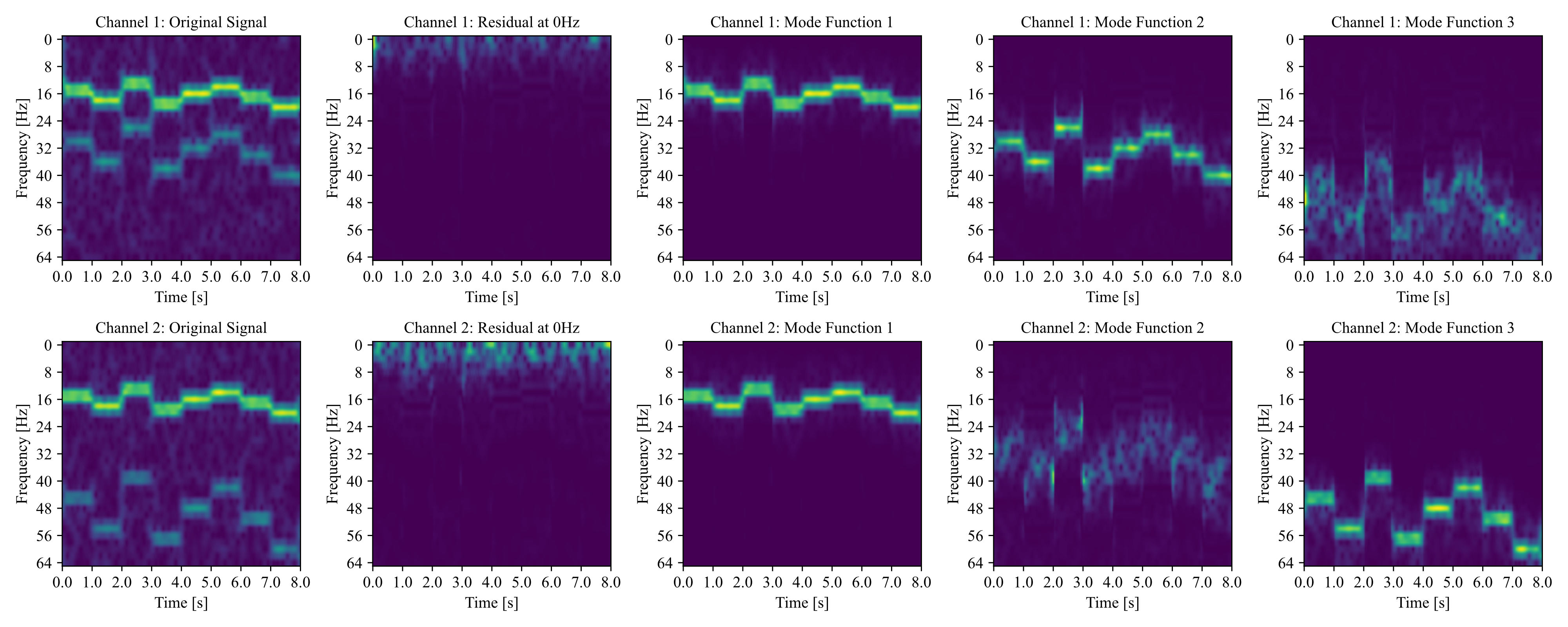}
    \label{fig:2.4a}
    }
    \subfigure[The time-varying central frequencies of the mode functions in the validation of mode alignment of dynamic STVMD. Left: central frequencies of the decomposed mode functions; Middle: the time-varying frequencies of the simulated signals in Channel 1; Right: the time-varying frequencies of the simulated signals in Channel 2. The frequency changes in the two-channel simulated signals are well tracked by the central frequencies of dynamic STVMD.]
    {
    \includegraphics[width=\textwidth]{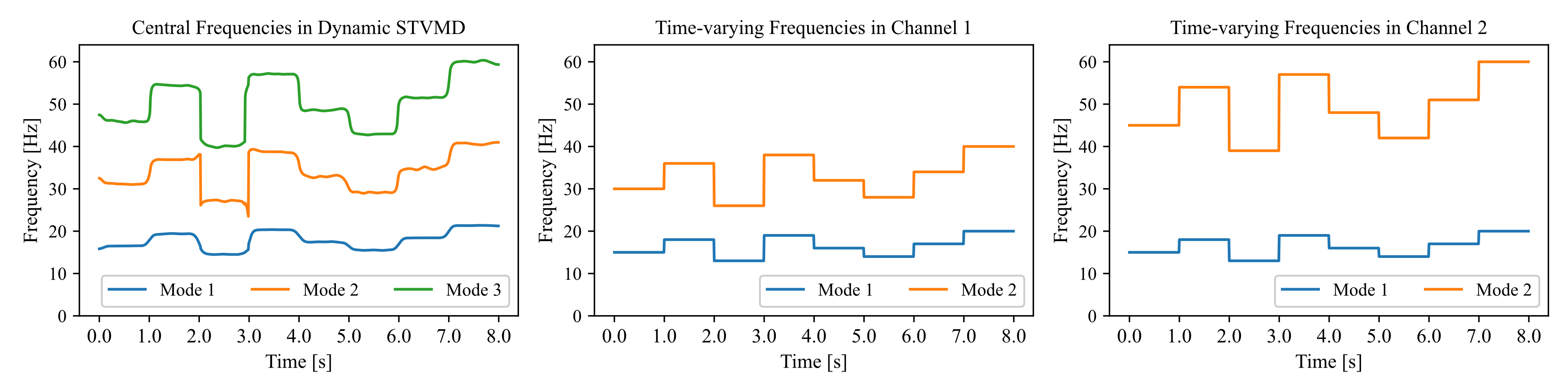}
    \label{fig:2.4b}
    }
\end{figure*}

\subsection{Real Application Scene}
Electroencephalogram (EEG) signal is a biosignal acquired from the scalp of human brain. The steady-state visual evoked potential (SSVEP) is a brain activtity which can be analyzed with EEG signals acquired from visual cortex \cite{nakanishi2018enhancing}. SSVEP refers to the brain signal evoked by external stimulus with certain frequencies. For example, when the subject views a stimulus at a frequency of 10$Hz$, the EEG signal in the visual cortex exhibits corresponding frequencies at 10 $Hz$ and its harmonics, such as 20 $Hz$ and 30 $Hz$, with decreased power.

In this experiment, the dynamic STVMD is applied to EEG signals in channel $O_z$, where $O_z$ is the electrode located in the visual cortex in the international 10-20 system. The signal is a 10-seconds time series with sampling rate 250 $Hz$. The first five seconds and the last five seconds of the EEG signals are acquired when the subject is stimulated with 10$Hz$ and 13 $Hz$ stimuli, respectively. The responses to 10 $Hz$ in the visual cortex are at 10 $Hz$ and it harmonics like 20 $Hz$ and 30 $Hz$. The responses to 13 $Hz$ in the visual cortex are at 13 $Hz$ and it harmonics like 26 $Hz$ and 39 $Hz$.

\begin{figure*}[htbp]
    \centering
    \includegraphics[width=\textwidth]{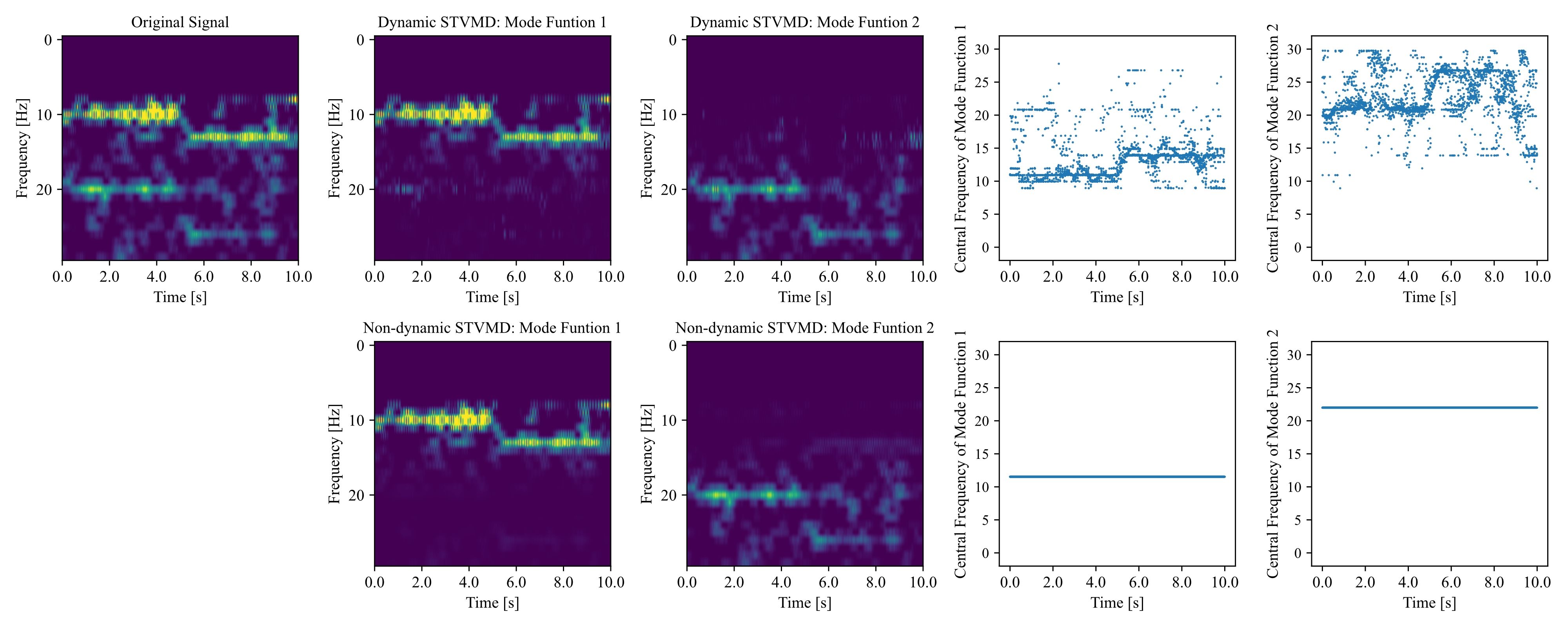}
    \caption{The decomposition results comparison between dynamic STVMD and non-dynamic STVMD in real SSVEP signals. The real signals are bandpassed between 5$Hz$ and 30$Hz$. The length of time window in STVMD is set to 250. The signals are decomposed into a residual at 0$Hz$ and two mode functions. We can find that the harmonic frequency changes in the real SSVEP signals are reflected in the central frequency changes of dynamic STVMD, while traditional VMD cannot.}
    \label{fig:2.5}
\end{figure*}

\begin{figure*}[htbp]
    \centering
    \includegraphics[width=\textwidth]{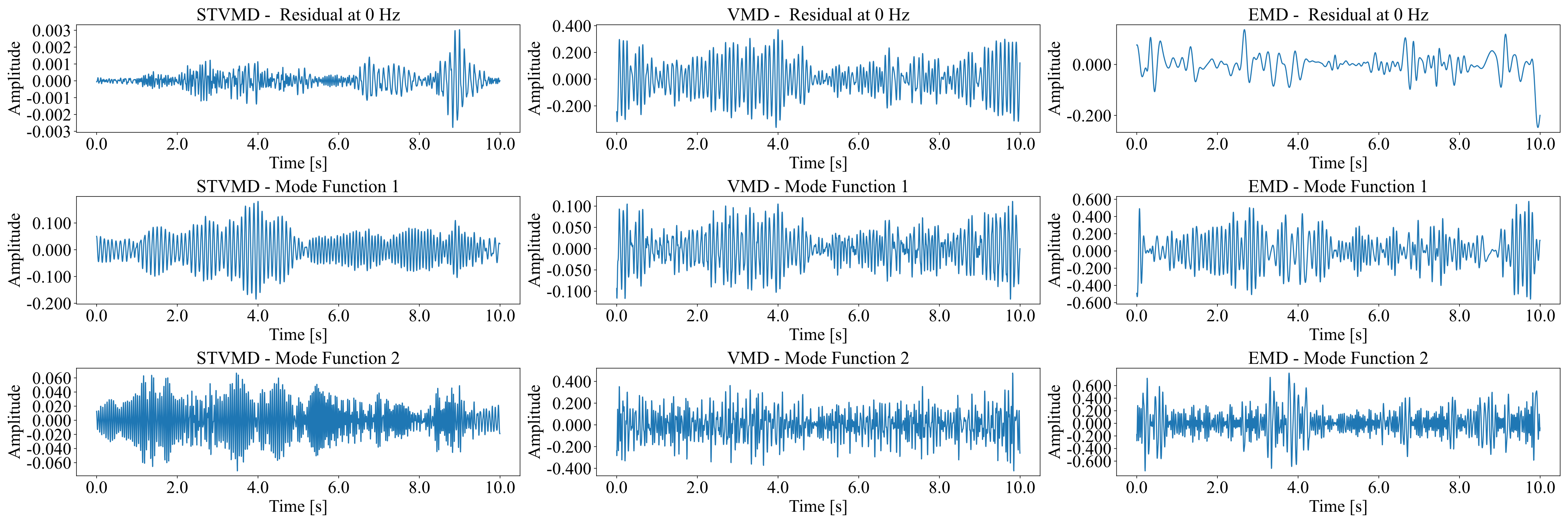}
    \caption{Comparison of decomposed mode functions among dynamic STVMD (ours), VMD\cite{dragomiretskiy_variational_2014} and EMD\cite{huang_empirical_1998} in real SSVEP signals. The dynamic STVMD holds the residual with lower amplitude than traditional VMD. It indicates that STVMD has smaller reconstruction error compared to VMD in the case of dynamic signal decomposition. The phenonmena in this figure is consistent with the results in Table \ref{tab:2}, where dynamic STVMD can reduce the reconstruction error.}
    \label{fig:2.6}
\end{figure*}

To remove the influence of noise, the signals are averaged from six epochs and then normalized. In the frequency domain, the power of signals in the low-frequency bands exhibits a decreasing trend. We first average the time-frequency map across the time dimension and obtain the power changes in the frequency domain. The trend is estimated using the five adjacent values (two to the left, two to the right and the same central value). To eliminate the influence of the trend, the trend is removed from the time-frequency map in STVMD. Taking into account that the power decreases with increasing frequencies, we use two components of the responses so that the input signals for STVMD are bandpassed between 5$Hz$ and 30$Hz$. The decomposition comparison between dynamic STVMD and non-dynamic STVMD are shown in Figure \ref{fig:2.5}. The comparison of mode functions between non-dynamic STVMD, VMD and EMD is given in Figure \ref{fig:2.6}. 


\section{Discussion}
In this study, we introduced the STVMD method for analyzing non-stationary signals, presenting both non-dynamic and dynamic variants that offer distinct advantages in signal decomposition. Our experimental results demonstrate key advancements in handling time-varying frequencies while revealing important insights about the method's capabilities and limitations.

The relationship between non-dynamic STVMD and traditional VMD/MVMD proves particularly interesting. As shown in our experimental results (Figure \ref{fig:1.1b} and \ref{fig:1.2c}), non-dynamic STVMD's performance converges to that of VMD/MVMD as the window length increases. This convergence suggests that non-dynamic STVMD effectively generalizes VMD/MVMD, with the window length parameter providing additional control over the time-frequency resolution trade-off. The key innovation lies in replacing the global Fourier transform with STFT, enabling localized frequency analysis while maintaining the theoretical foundations of variational mode decomposition.

Dynamic STVMD represents a more fundamental advancement in handling non-stationary signals. The time-varying central frequencies provide a crucial advantage in tracking dynamic frequency components, as evidenced by the significantly lower reconstruction errors shown in Table \ref{tab:2}. Our experiments with simulated non-stationary signals demonstrate that dynamic STVMD achieves superior mode separation compared to both VMD and non-dynamic STVMD. This improvement stems from the method's ability to adapt its frequency bands according to local signal characteristics, rather than enforcing fixed frequency partitions across the entire signal duration.

The application to real EEG signals reveals practical benefits of dynamic STVMD in biomedical signal processing. The method successfully tracked frequency transitions in steady-state visual-evoked potentials, demonstrating its utility in analyzing brain signals where frequency components evolve over time. This capability has significant implications for brain-computer interface applications and neurological diagnostics.

However, our analysis also reveals important limitations. The computational complexity increases with the number of mode functions, potentially limiting real-time applications. Additionally, the inherent trade-off between time and frequency resolution in STFT carries over to STVMD, requiring careful parameter selection based on application requirements. Small time windows may lead to poor frequency resolution, while large windows can miss rapid frequency changes and introduce latency in online processing.

Looking forward, several promising research directions emerge. First, adaptive parameter selection mechanisms could optimize the window length and number of modes based on signal characteristics. Second, the integration of machine learning techniques might enable more efficient computation and automated parameter tuning. Finally, the theoretical framework of STVMD could be extended to handle multivariate signals with more complex inter-channel relationships.

\bibliographystyle{unsrt}
\bibliography{main}

\end{document}